\pdfoutput=1
\documentclass[12pt]{article}
\usepackage{fancyhdr}
\usepackage{datetime} 

\usepackage{amsmath,amssymb,amsbsy,exscale}
\usepackage{graphicx}
\usepackage{epsfig}
\usepackage{multicol}
\usepackage{mathrsfs}
\usepackage{mathtools}
\usepackage{blindtext}
 \usepackage{fancyhdr}
\usepackage{hyperref}
\usepackage{cite}

\usepackage{enumerate}

\usepackage{dsfont}
\usepackage{appendix}

\usepackage{parskip}

\usepackage{xspace}
\usepackage{braket}
\usepackage{array}
\usepackage{feynmp}
\usepackage{pinlabel}
\usepackage[top=1.2in, bottom=1.2in,left=0.8in, right=0.8in,includefoot]{geometry}          

\usepackage{graphicx,rotating,colortbl}
\definecolor{nicered}{rgb}{0.7,0.1,0.1}
\definecolor{nicegreen}{rgb}{0.1,0.5,0.1}

\definecolor{rosso}{cmyk}{0,1,1,0.4}
\definecolor{babypink}{rgb}{0.96, 0.76, 0.76}
\definecolor{babyblueeyes}{rgb}{0.63, 0.79, 0.95}
\definecolor{azure(colorwheel)}{rgb}{0.0, 0.5, 1.0}
\definecolor{amethyst}{rgb}{0.6, 0.4, 0.8}
\definecolor{MyDarkBlue}{rgb}{0,0.1,0.7}

\definecolor{secnum}{RGB}{13,151,225}
\definecolor{ptcbackground}{RGB}{212,237,252}
\definecolor{ptctitle}{RGB}{0,177,235}
\definecolor{blus}{cmyk}{1,1,0,0.1}
\definecolor{verdes}{cmyk}{0.99,0,0.59,0.65}
\definecolor{rossos}{cmyk}{0,1,1,0.55}
\definecolor{redy}{cmyk}{0,1,1,0.7}
\definecolor{greeny}{cmyk}{0.99,0,0.59,0.98}
\definecolor{green-go}{cmyk}{0.79,0,0.59,0.5}

\usepackage{hyperref, pdfsync, epstopdf, enumerate}
\hypersetup{colorlinks,bookmarksopen,bookmarksnumbered,citecolor={nicegreen},
linkcolor={MyDarkBlue},pdfstartview=FitH,urlcolor={amethyst}}
\usepackage{amsfonts}
\usepackage{slashed}

\usepackage{verbatim}
\usepackage{doi}
\usepackage{url}
\usepackage{array}

\usepackage{cite}
\usepackage{float}
\usepackage[labelfont=bf]{caption}
\usepackage{subcaption}
\usepackage{parskip}
\usepackage{hhline}

\numberwithin{equation}{section}

\usepackage{sectsty}

\usepackage{mdframed}
\usepackage{titletoc}

\titlecontents{lsection}
  [5.8em]{\sffamily}
  {\color{greeny}\contentslabel{2.3em}\normalcolor}{}
  {\titlerule*[1000pc]{.}\contentspage\\\hspace*{-5.8em}\vspace*{5pt}%
    \color{white}\rule{\dimexpr\textwidth-15.5pt\relax}{1pt}}

\usepackage{titlesec}
\titlelabel{\thetitle.\quad}

\newcommand{\tmtextbf}[1]{{\bfseries{#1}}}
\newcommand{\tmtextrm}[1]{{\rmfamily{#1}}}

\newcommand{\gappeq}{{\rlap{{\raise}.5ex\text{\ensuremath{>}}}{{\lower}.5ex\text{\ensuremath{\sim}}}}}
\newcommand{\lappeq}{{\rlap{{\raise}.5ex\text{\ensuremath{<}}}{{\lower}.5ex\text{\ensuremath{\sim}}}}}

\newcommand{\I}{\tmtextrm{1{\kern}-.24em l}}

\newcommand{\newc}{\newcommand}
\newc{\be}{\begin{equation}}
\newc{\ee}{\end{equation}}
\newc{\bal}{\begin{align}}
\newc{\eal}{\end{align}}
\newc{\ba}{\begin{eqnarray}}
\newc{\ea}{\end{eqnarray}}
\newc{\bea}{\begin{eqnarray*}}
\newc{\eea}{\end{eqnarray*}}
\newc{\D}{\partial}
\newc{\som}{\sin\omega}
\newc{\com}{\cos\omega}
\newc{\sth}{\sin\theta}
\newc{\cth}{\cos\theta}
\newc{\stom}{\sin^2\omega}
\newc{\ctom}{\cos^2\omega}
\newc{\stth}{\sin^2\theta}
\newc{\ctth}{\cos^2\theta}
\newc{\ie}{{\it i.e.} }
\newc{\eg}{{\it e.g.} }
\newc{\etc}{{\it etc.} }
\newc{\etal}{{\it et al.}}

\def\beq{\begin{equation}}
\def\eeq{\end{equation}}
\def\bea{\begin{eqnarray}}
\def\eea{\end{eqnarray}}

\let\eps=\epsilon
\let\lam=\lambda
\let\kap=\kappa
\let\alp=\alpha

\newcommand{\dd}{\text{d}}

\newcommand{\lapproxeq}{\lower .7ex\hbox{$\;\stackrel{\textstyle
<}{\sim}\;$}}
\newcommand{\gapproxeq}{\lower .7ex\hbox{$\;\stackrel{\textstyle
>}{\sim}\;$}}
\newcommand{\stackdown}[2]{\lower 1.4ex\hbox{$\;\stackrel{\textstyle{#1}}
{\scriptstyle{#2}}\;$}}

\newcommand{\Inm}{\mathcal{I}_m}
\newcommand{\Iv}{\mathcal{I}_{\mathcal{V}}}
\newcommand{\Iphi}{\mathcal{I}_\phi}
\newcommand{\hatt}{\hat{t}}
\newcommand{\bart}{\bar{t}}
\newcommand{\hk}{\hat{\kappa}}
\newcommand{\bk}{\bar{\kappa}}

\newcommand{\bl}{\bar{\lambda}}
\newcommand{\heps}{\hat{\epsilon}}
\newcommand{\beps}{\bar{\epsilon}}
\newcommand{\hH}{\hat{H}}
\newcommand{\bH}{\bar{H}}

\makeatletter
\def\@xfootnote[#1]{%
  \protected@xdef\@thefnmark{#1}%
  \@footnotemark\@footnotetext}
\makeatother

\numberwithin{equation}{section}

\allowdisplaybreaks

\begin{document}

\topmargin -1.0cm
\oddsidemargin -0.5cm
\evensidemargin -0.5cm

 
{\vspace{1cm}}
\begin{center}
\vspace{1cm}

 {\LARGE  \tmtextbf{ 
{Frame-dependence of higher-order inflationary observables in scalar-tensor theories}
}} {\vspace{.5cm}}\\

\vspace{1.4cm}

{\large  {\bf Alexandros Karam\footnote[1]{{\href{mailto:alkaram@cc.uoi.gr}{alkaram@cc.uoi.gr}}}, Thomas Pappas\footnote[2]{{\href{mailto:thpap@cc.uoi.gr}{thpap@cc.uoi.gr}}} and Kyriakos Tamvakis\footnote[3]{{\href{mailto:tamvakis@uoi.gr}{tamvakis@uoi.gr}}}
}
\vspace{.3cm}

{\it }\

{\em  \normalsize 

~Department of Physics, University of Ioannina, GR--45110 Ioannina, Greece


}
\vspace{0.5cm}

}
\vspace{1.2cm}
 \end{center}
\noindent --------------------------------------------------------------------------------------------------------------------------------

\vspace{-0.3cm}

\begin{abstract}

\noindent {\normalsize In the context of scalar-tensor theories of gravity we compute the third-order corrected spectral indices in the slow-roll approximation. The calculation is carried out by employing the Green's function method for scalar and tensor perturbations in both the Einstein and Jordan frames. Then, using the interrelations between the Hubble slow-roll parameters in the two frames we find that the frames are equivalent up to third order.
Since the Hubble slow-roll parameters are related to the potential slow-roll parameters, we express the observables in terms of the latter which are manifestly invariant. Nevertheless, the same inflaton excursion leads to different predictions in the two frames since the definition of the number of e-folds differs. To illustrate this effect we consider a nonminimal inflationary model and find that the difference in the predictions grows with the nonminimal coupling and it can actually be larger than the difference between the first and third order results for the observables. Finally, we demonstrate the effect of various end-of-inflation conditions on the observables. These effects will become important for the analyses of inflationary models in view of the improved sensitivity of future experiments.}

\end{abstract}

\vspace{.1cm}

\noindent --------------------------------------------------------------------------------------------------------------------------------


\vspace{0.2cm}


\vspace{0.6cm}


{
\tableofcontents
}

\vspace{.9cm}
\noindent --------------------------------------------------------------------------------------------------------------------------------
\newpage
\section{Introduction}
\label{sec:intro}

The theory of cosmic inflation was originally advocated as a solution to the flatness and horizon problems \cite{Guth1981, Linde1982} of the big-bang cosmology. When treated quantum-mechanically, inflation can also provide a mechanism for the generation of the perturbations that have resulted in the anisotropies observed in the cosmic microwave background \cite{Hawking1982, Starobinsky1982, Guth1982, Linde1983a}. It is usually formulated in terms of a single scalar field, minimally coupled to gravity, whose potential energy dominates over its kinetic energy for a short period of time and drives the accelerated expansion of the universe. This phase can be most easily achieved if the scalar potential $\mathcal{V}(\phi)$ has a relatively flat plateau and the scalar field can slowly roll down until it reaches the minimum. 

Over the years a vast plethora of inflationary models have been proposed, originating from diverse physics frameworks. Recently, the increasing sensitivity of the experiments, and in particular measurements from the Planck and BICEP2/Keck Collaborations \cite{Ade2015, Ade2016}, have put stringent constraints on many of these models. The simplest models, where a single scalar field is minimally coupled to gravity, seem to be disfavored\footnote{See however \cite{Ballesteros2016}.}. On the other hand, slightly more convoluted models such as the Starobinsky model \cite{Starobinsky1980, Pallis2017, Bruck2015, Copeland2015, Kehagias2014, Asaka2016}, nonminimal Higgs inflation \cite{Bezrukov2008, Bezrukov2009, DeSimone2009, Barbon2009, Bezrukov2009a, Barvinsky2009, Lerner2010, Lerner2010a, Popa2011, Bezrukov2011, Kamada2012, Steinwachs2012, Bezrukov2013, George2014, Hamada2014, Hamada2014a, Hamada2015, Bezrukov2014, Allison2014, Salvio2015, Bruck2016, Calmet2016}, or the so-called $\alpha$--attractors \cite{Kallosh2013b, Kallosh2013d, Kallosh2013a, Kallosh2014a, Linde2015, Galante2015, Broy2015, Kallosh2015, Carrasco2015, Yi2016, Odintsov2016, Bhattacharya2017, Dimopoulos2017} give predictions for the observables that lie inside the sweet spot of the measurements. A common feature of these models is that they can be formulated in terms of a nonminimal coupling function $F(\phi)$ between the inflaton $\phi$ and the scalar curvature $R$. Such nonminimal coupling is expected to be generated at the quantum level of the theory even if it is absent in the classical action \cite{Faraoni1996}. These nonminimally coupled theories belong to a general class of gravity theories termed \textit{scalar-tensor (ST) theories} \cite{Faraoni2004a}. Other examples of such theories include, among others, the $f(R)$ models \cite{Sotiriou2010, Capozziello2011, Nojiri2011, DeFelice2010, Clifton2012, Rinaldi2014, Bamba2014, Nojiri2017}, scale-invariant models \cite{Shaposhnikov2009a, Khoze2013, Kannike2014, Gabrielli2014, Salvio2014, Csaki2014a, Kannike2015, Barrie2016, Marzola2016a, Marzola2016b, Rinaldi2016a, Farzinnia2016, Kannike2016, Rinaldi2016b, Karananas2016, Tambalo2017, Kannike2017, Ferreira2017, Salvio2017, Kannike2017a} and nonminimal inflationary models \cite{Fakir1990, Makino1991, Faraoni1996, Torres1997, Faraoni2000, Koh2005, Park2008a, Nozari2008, Bauer2008, Nozari2010, Okada2010, Pallis2010, Edwards2014, Kehagias2014, Inagaki2015, Artymowski2017}.

Scalar-tensor theories are usually formulated in either the \textit{Jordan frame (JF)} or the \textit{Einstein frame (EF)}. In the JF the Planck mass is a dynamical quantity that depends on the value of the scalar field, whose self-interactions are described by a potential. Furthermore, the scalar field is minimally coupled to the metric, and the matter part of the action is just the standard one. In the EF the gravitational action has the standard Einstein-Hilbert form plus a scalar field described by an effective potential. Moreover, the scalar appears in the matter sector of the action through the rescaling factor which multiplies the metric tensor. The two frames are mathematically equivalent at the classical level\footnote{See also \cite{Kamenshchik2015, Herrero-Valea2016, Pandey2016, Pandey2017} for considerations on the quantum equivalence of the frames.} since one can always switch between them by applying a conformal transformation of the metric and a field redefinition, collectively referred to as \textit{frame transformation}. Nevertheless, the physical equivalence of the frames with respect to the physical predictions has become a matter of a long-standing debate \cite{Capozziello1997, Dick1998, Faraoni1999, Faraoni1999a, Flanagan2004, BHADRA2007, Nozari2009, Capozziello2010, Corda2011, Qiu2012, Qiu2012a, Quiros2013, Chiba2013a, Postma2014, Qiu2015, Domenech2015a, Bahamonde2016, Brooker2016, Bhattacharya2017a, Bahamonde2017}.

Inflation is usually studied with the help of the so-called \textit{slow-roll parameters} which are generally frame-covariant \cite{Kaiser1995, Nozari2010, DeFelice2011, Burns2016, Karamitsos2017}. Nevertheless, if we analyze the slow-roll regimes in the JF and EF using invariant quantities then we can quickly move between different parametrizations. This invariant formalism was recently proposed and developed in \cite{Jaerv2015, Jaerv2015a, Kuusk2016, Kuusk2016a, Jaerv2017}. In \cite{Kuusk2016} the authors calculated the spectral indices up to second order in the slow-roll parameters in both the EF and JF and showed that the two frames are physically equivalent. Here we extend their results up to third order in the slow-roll parameters and also examine how the different definitions for the number of $e$-folds in the two frames affect the observables.

This paper is organized as follows: in Sec. \ref{sec:formalism} we review the invariant formalism introduced in \cite{Kuusk2016}. After presenting the three principal quantities which are invariant under a conformal transformation of the metric and a redefinition of the scalar field, we consider the slow-roll approximation in the two frames and define the corresponding \textit{Hubble slow-roll} parameters (HSRPs). We also define a hierarchy of \textit{potential slow-roll} parameters (PSRPs) which are frame independent. As shown in \cite{Jaerv2017}, this formalism proves to be attractive since many inflationary models can be classified according to the form of their invariant potentials. This provides an elegant explanation as to why vastly different models can produce the same predictions for the inflationary observables. 

In Sec. \ref{sec:spectra-indices} we adopt the Green's function method considered in \cite{Gong2001} and calculate the spectral indices up to third order in the slow-roll parameters in both the JF and EF. Then, using the relations between the HSRPs we find that the two frames are equivalent. Furthermore, since the HSRPs can be related to the PSRPs, we express the spectral indices in terms of the PSRPs which are manifestly frame invariant. 

In Sec. \ref{sec:efolds} we consider the nonminimal Coleman-Weinberg model developed in \cite{Kannike2016} and compare the predictions of the third order corrected expressions we obtained with the most commonly used first order results. Furthermore, even though the expressions for the observables that we obtain are frame invariant, the definition of the number of $e$-folds is not and this results to different predictions. To this end, we examine how the predictions change if the required 50--60 number of $e$-folds is taken in the Einstein or in the Jordan frame. Finally, we examine how the predicted values for the inflationary observables are affected by the end-of-inflation condition. The \textit{exact} condition for inflation to end is when $\eps_H=1$. The usual approach is to Taylor approximate this condition with PSRPs. Most authors use only the first order approximation $\eps_H \approx \eps_V$ since this is indeed a good approximation for almost all of the inflationary epoch save for the last few $e$-folds before inflation ends when this approximation breaks down. Since we have obtained the third-order corrected expressions for the inflationary observables we also compare the results against three more end-of-inflation conditions, namely, the third-order Taylor approximation of the condition $\eps_H=1$ with PSRPs, as well as against the Pad\'{e} [1/1] and Pad\'{e} [2/2] approximants. All of these considerations prove to be relevant since the differences in the predictions that we obtain are within the accuracy of future experiments and may prove instrumental in ruling out various inflationary models.

In Sec. \ref{sec:Conclusions} we summarize our results and conclude. Useful formulas are presented in the Appendixes. 

\section{Invariant formalism and slow-roll approximation}
\label{sec:formalism}

In this section, we consider the general action of a single scalar field that describes a wide class of scalar-tensor gravity theories. By using the frame and parametrization invariant formalism introduced in \cite{Jaerv2017, Jaerv2015, Jaerv2015a, Kuusk2016a, Kuusk2016} we write down the field equations of motion in terms of quantities that are invariant under conformal rescalings of the metric and redefinitions of the scalar field.

\subsection{General action}
\label{subsec:action}

The most general action for scalar-tensor theories has the form \cite{Flanagan2004}
\be 
S = \int \dd^4 x \sqrt{-g} \left\lbrace \frac{1}{2}\mathcal{A}(\Phi) R - \frac{1}{2}B(\Phi) g^{\mu\nu} \left( \nabla_\mu \Phi \right) \left( \nabla_\nu \Phi \right) - \mathcal{V}(\Phi) \right\rbrace + S_m \left[ e^{2 \sigma(\Phi)} g_{\mu\nu} , \chi \right] ,
\label{Action}
\ee
where in the first term $g$ is the metric determinant, $R$ denotes the Ricci scalar associated with the metric $g_{\mu\nu}$ and $\mathcal{V} (\Phi)$ is the scalar potential. In the second term, $S_m$ stands for the matter part of the action. Furthermore, the four functions $\mathcal{A} (\Phi)$, $\mathcal{B} (\Phi)$, $\mathcal{V} (\Phi)$ and $\sigma (\Phi)$ are arbitrary dimensionless functions of the scalar field $\Phi$ that completely characterize a model, and we call them \textit{model functions}. Throughout, we normalize $\Phi$ in terms of the reduced Planck mass, $M_P / (8 \pi G)^{1/2} \equiv 1$.

We assume that the background metric is the flat Friedmann--Lema\^{i}tre--Robertson--Walker (FLRW) with the space-positive signature
\be
ds^2= a^2(\tau)(-d\tau^2+dx^2+dy^2+dz^2) ,
\label{FLRW_metric}
\ee
where $a(\tau)$ is the scale factor of the Universe as a function of the frame-invariant conformal time.
By considering a rescaling of the metric
\be 
g_{\mu\nu} = e^{2 \bar{\gamma} (\bar{\Phi})} \bar{g}_{\mu\nu}
\label{trans_metric1}
\ee
and a redefinition of the field
\be 
\Phi = \bar{f} (\bar{\Phi})
\label{trans_field}
\ee
one can easily verify that the action \eqref{Action} is invariant up to a boundary term, if the model functions transform according to the following relations:
\ba 
\bar{\mathcal{A}}(\bar{\Phi}) &=& e^{2 \bar{\gamma} (\bar{\Phi})} \mathcal{A} \left( \bar{f} (\bar{\Phi}) \right) ,
\label{trans_A}
\\
\nonumber \\
\bar{\mathcal{B}}(\bar{\Phi}) &=& e^{2 \bar{\gamma} (\bar{\Phi})} \left[ (\bar{f}')^2 \mathcal{B} \left( \bar{f} (\bar{\Phi}) \right) - 6 (\bar{\gamma}')^2 \mathcal{A} \left( \bar{f} (\bar{\Phi}) \right) - 6 \bar{\gamma}' \bar{f}' \mathcal{A}'  \right] ,
\label{trans_B}
\\
\nonumber \\
\bar{\mathcal{V}} (\bar{\Phi}) &=& e^{4 \bar{\gamma} (\bar{\Phi})} \mathcal{V} \left( \bar{f} (\bar{\Phi}) \right) ,
\label{trans_V}
\\ 
\nonumber \\
\bar{\sigma} (\bar{\Phi}) &=& \sigma \left( \bar{f} (\bar{\Phi})  \right) + \bar{\gamma} (\bar{\Phi}) ,
\label{trans_sigma}
\ea
where a prime indicates differentiation with respect to the argument, e.g. $\bar{\gamma}' \equiv \dd \bar{\gamma} (\bar{\Phi}) / \dd \bar{\Phi}$ and $\mathcal{A}' \equiv \dd \mathcal{A}(\Phi) / \dd \Phi$, and an overbar denotes quantities which are given in terms of the conformal metric $\bar{g}_{\mu\nu}$.

Now, using the transformations \eqref{trans_metric1}-\eqref{trans_field} one can fix two out of the four arbitrary functions $\left\lbrace \mathcal{A} , \mathcal{B} , \mathcal{V} , \sigma \right\rbrace $. Different choices for these functions correspond to different \textit{parametrizations}. For example, the choice 
\be
\mathcal{A} = F(\phi), \quad \mathcal{B} = 1 , \quad \mathcal{V} = \mathcal{V} (\phi) , \quad \sigma = 0 ,
\ee
corresponds to the JF in the Boisseau-Esposito-Far\`{e}se-Polarski-Starobinski parametrization \cite{Boisseau2000, Esposito-Farese2001}, the choice 
\be 
\mathcal{A} = \Psi, \quad \mathcal{B} = \frac{\omega(\Psi)}{\Psi} , \quad \mathcal{V} = \mathcal{V} (\Psi) , \quad \sigma = 0  ,
\ee
corresponds to the JF in the Brans-Dicke-Bergmann-Wagoner parametrization \cite{Brans1961, Bergmann1968, Wagoner1970}, while the choice
\be 
\mathcal{A} = 1, \quad \mathcal{B} = 2 , \quad \mathcal{V} = \mathcal{V} (\varphi) , \quad \sigma = \sigma(\varphi)  ,
\ee
represents the EF in the canonical parametrization \cite{Dicke1962a, Brans1961, Bergmann1968, Wagoner1970}.

\subsection{Invariants}
\label{subsec:invariants}

Next, we follow \cite{Kuusk2016} and consider three quantities which are invariant under a conformal rescaling of the metric and a reparametrization of the scalar field as a result of the transformation properties \eqref{trans_A}-\eqref{trans_sigma} of the model functions. These invariants are
\be 
\Inm (\Phi) \equiv \frac{e^{2 \sigma (\Phi)}}{\mathcal{A} (\Phi)} ,
\label{inv_m}
\ee
\be 
\Iv (\Phi) \equiv \frac{\mathcal{V} (\Phi)}{(\mathcal{A} (\Phi))^2} ,
\label{inv_v}
\ee
\be 
\Iphi (\Phi) \equiv \int \left( \frac{2 \mathcal{A} \mathcal{B} + 3 (\mathcal{A}')^2}{4 \mathcal{A}^2} \right)^{1/2} \dd \Phi .
\label{inv_phi}
\ee
The first invariant, $\Inm(\Phi)$, is a quantity that characterizes the nonminimality of a theory. For constant $\Inm(\Phi)$ the scalar field is minimally coupled to gravity, and we are dealing with standard general relativity. On the other hand, if $\Inm'(\Phi) \not\equiv 0$, then this invariant is a dynamical function and the scalar field is nonminimally coupled to gravity, as is the case in the JF. The second invariant, $\Iv(\Phi)$, contains the self-interactions of the scalar field and plays the role of an invariant potential. Finally, the third invariant, $\Iphi(\Phi)$, measures the volume of the one-dimensional space of the scalar field and can be interpreted as the invariant propagating scalar degree of freedom. 

The transformation properties of the model functions can also be used to define tensorial invariants, for example~\cite{Kuusk2016}
\be 
\hat{g}_{\mu\nu} \equiv \mathcal{A}(\Phi) g_{\mu\nu} .
\label{trans_metric2}
\ee
The above choice is not unique since the tensor \eqref{trans_metric2} does not change its transformation properties if it is multiplied by a scalar invariant, i.e.,
\be 
\bar{g}_{\mu\nu} \equiv e^{2 \sigma (\Phi)} g_{\mu\nu} = \Inm \hat{g}_{\mu\nu}
\label{trans_metric3}
\ee
is also invariant under the transformations \eqref{trans_metric1} and \eqref{trans_field}. 

In the following, a barred or a hatted variable will represent the quantity evaluated in the JF or EF, respectively. The relation between the time coordinate, the scale factor and the Hubble parameter in the two frames is \cite{Kuusk2016}
\be 
\frac{\dd}{\dd \bar{t}} = \frac{1}{\sqrt{\Inm}} \frac{\dd}{\dd \hat{t}} , \quad \bar{a} (\bar{t}) = \sqrt{\Inm} \hat{a} (\hat{t}) ,
\label{trans_time_scale-factor}
\ee
\be 
\bH = \frac{1}{\sqrt{\Inm}} \left( \hH + \frac{1}{2} \frac{\dd \ln{\Inm}}{\dd \hatt} \right) .
\label{trans_Hubble}
\ee

\begin{figure}
\includegraphics[width=.5\textwidth]{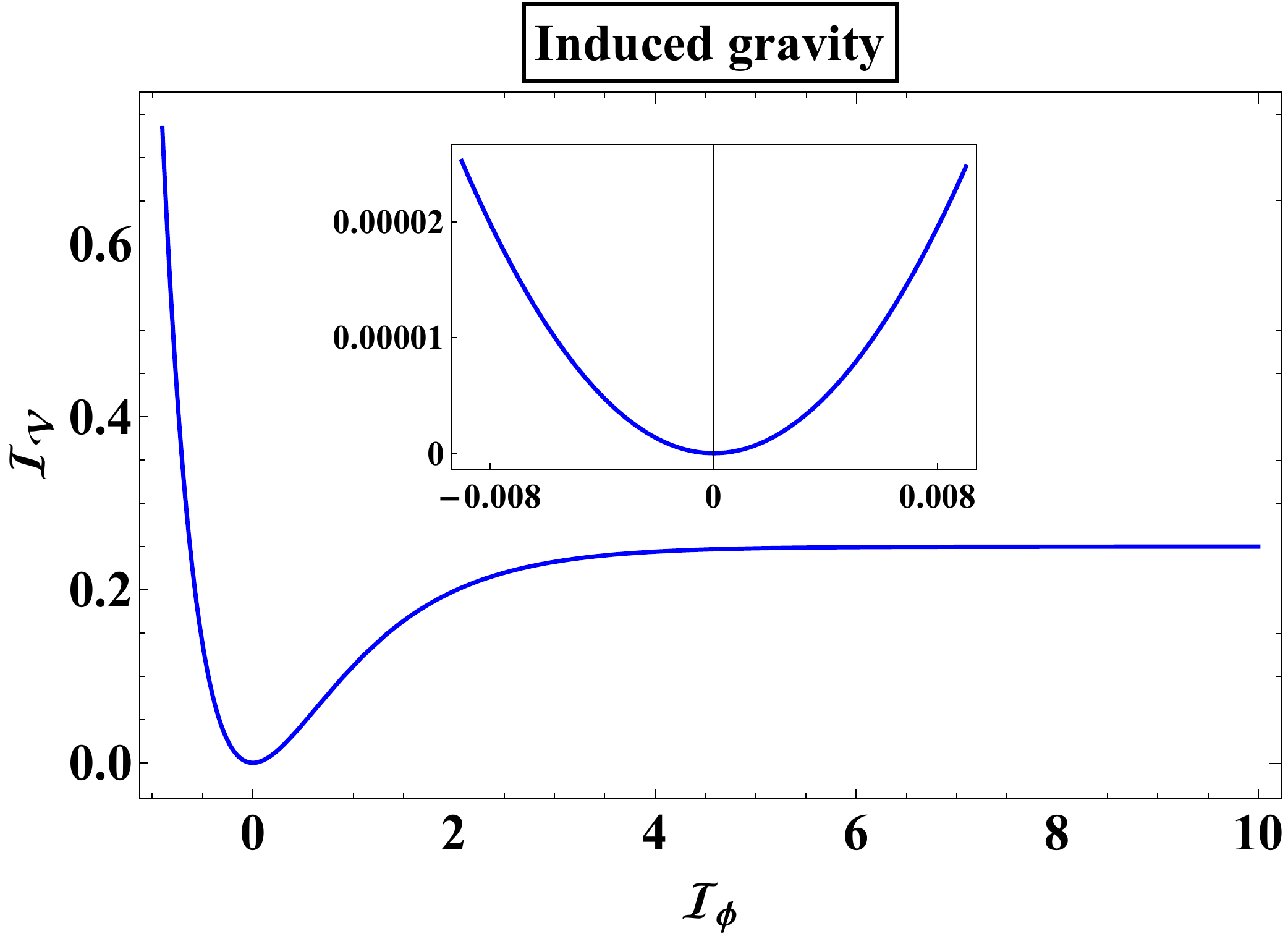}
\includegraphics[width=.5\textwidth]{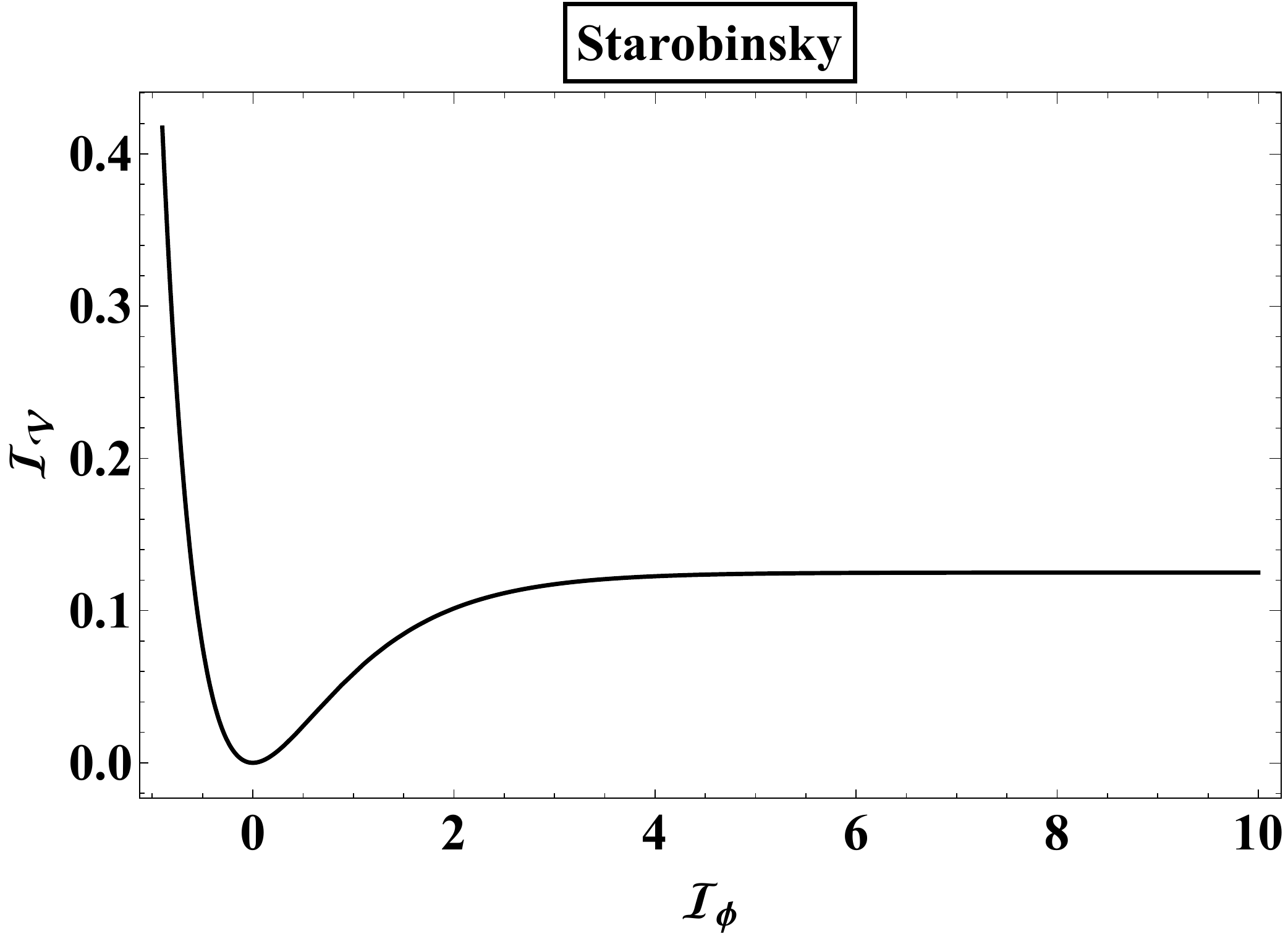}
\caption{The normalized invariant inflationary potentials for induced gravity and Starobinsky models for $\xi=2$. In the strong coupling limit the invariant potentials have a similar form and lead to the same predictions, while in the limit \eqref{Induced_to_quadratic_condition} induced gravity approaches the quadratic inflation attractor (inset in left plot).}
\label{fig:Induced-v-Staro}
\end{figure}

An interesting and appealing feature of the invariant formalism, which was pointed out in \cite{Jaerv2017}, is that inflationary models with very different background physical motivations can be described by similar invariant potentials and thus lead to the same predictions for the inflationary observables. As an example, let us consider \textit{induced gravity} inflation \cite{Accetta1985, Carugno1993, Kaiser1994, Kaiser1994a, Cervantes-Cota1995, Cerioni2009, Giudice2014} and \textit{Starobinsky} inflation \cite{Starobinsky1980, Pallis2017, Bruck2016a, Bruck2015, Copeland2015, Kehagias2014, Asaka2016}. The former is described by the model functions
\ba 
\mathcal{A} (\Phi) &=& \xi \Phi^2 ,
\\ 
\mathcal{B} (\Phi) &=& 1 ,
\\ 
\sigma (\Phi) &=& 0 ,
\\ 
\mathcal{V} (\Phi) &=& \lambda \left( \Phi^2 - v^2 \right)^2 ,
\ea
where $\xi$ is the nonminimal coupling and $v$ is the vacuum expectation value (VEV) of the scalar field $\Phi$ which induces the Planck mass scale,
\be 
1 = \xi v^2 .
\ee
For Starobinsky inflation with $f(R) = R + b R^2$ one has~\cite{Burns2016}
\ba 
\mathcal{A} (\Phi) &=& \Phi ,
\\ 
\mathcal{B} (\Phi) &=& 0 ,
\\ 
\sigma (\Phi) &=& 0 ,
\\ 
\mathcal{V} (\Phi) &=& \frac{b}{2} \left( \frac{\Phi-1}{2b}\right)^2.
\ea
Next, following the recipe of \cite{Jaerv2017} we can obtain the invariant potentials $\Iv$ for the two models. As a first step, using \eqref{inv_phi} we calculate the form of the invariant fields
\ba 
\text{Induced gravity:}&& \qquad \Iphi = \sqrt{\frac{1 + 6 \xi}{2 \xi}} \ln \left( \frac{\Phi}{v_\Phi} \right) ,
\\
\nonumber \\
\text{Starobinsky:}&& \qquad \Iphi = \frac{\sqrt{3}}{2} \ln \Phi .
\ea
Afterwards, inverting the above relations we find $\Phi(\Iphi)$ and then using \eqref{inv_v} we calculate \\
$\Iv(\Phi(\Iphi)) = \Iv(\Iphi)$ and obtain
\ba
\text{Induced gravity:}&& \qquad \Iv (\Iphi)= \frac{\lam}{\xi^2}\left( 1- e^{-\sqrt{\frac{8 \xi}{1+6 \xi}}\Iphi} \right)^2,
\label{Induced_Inv_pot}
\\
\nonumber \\
\text{Starobinsky:}&& \qquad \Iv (\Iphi)=\frac{1}{8 \, b} \left( 1- e^{-\frac{2}{\sqrt{3}}\Iphi}\right)^2.
\label{Staro_Inv_pot}
\ea
The forms of the invariant potentials suggest that for large values of the nonminimal coupling ($\xi \gtrsim 1$) the shape of the induced gravity invariant potential \eqref{Induced_Inv_pot} coincides with its Starobinsky counterpart \eqref{Staro_Inv_pot}, a behavior depicted in Fig.~\ref{fig:Induced-v-Staro}. As a consequence, the two models yield identical predictions in the strong coupling regime. On the other hand, in the weak coupling limit induced gravity gives the same predictions with quadratic inflation \cite{Linde1983a}. Indeed, when
\be
\Iphi \ll \sqrt{\frac{1+6 \xi}{8 \xi}} \, ,
\label{Induced_to_quadratic_condition}
\ee
the invariant potential for induced gravity becomes \cite{Kallosh2014, Kallosh2013a}
\be 
\Iv = M^2 \Iphi^2 , \quad \text{with} \quad M^2 = \frac{8 \lambda}{\xi \left( 1 + 6 \xi \right)} \, .
\ee
Note in \eqref{Induced_to_quadratic_condition} that as $\xi$ becomes smaller the allowed range for the field $\Iphi$ in which induced gravity and quadratic inflation produce similar predictions becomes wider. As a consequence, only for small values of $\xi$ the field $\Iphi$ can produce the required 50-60 number of $e$-folds. This is why the induced gravity predictions reach the quadratic inflation attractor in the small coupling regime.

\subsection{Slow-roll in the Jordan frame}
\label{subsec:SRJF}

Let us consider the slow rolling of the inflaton field in the JF. Taking the functional derivative of the action \eqref{Action} with respect to the metric and the scalar field in the JF, we can write down the equations of motion in terms of the invariants as
\ba 
\bH^2 &=& \frac{1}{3} \left( \frac{\dd \Iphi}{\dd \bart} \right)^2 + \bH \frac{\dd \ln{\Inm}}{\dd \bart} - \frac{1}{4} \left( \frac{\dd \ln{\Inm}}{\dd \bart} \right)^2 + \frac{1}{3} \frac{\Iv}{\Inm} ,
\label{EOM1A_JF}
\\ 
\nonumber \\
\frac{\dd \bH}{\dd \bart} &=& - \frac{1}{2} \bH \frac{\dd \ln{\Inm}}{\dd \bart} + \frac{1}{4} \left( \frac{\dd \ln{\Inm}}{\dd \bart} \right)^2 - \left( \frac{\dd \Iphi}{\dd \bart} \right)^2 + \frac{1}{2} \frac{\dd^2 \ln{\Inm}}{\dd \bart^2} ,
\label{EOM2A_JF}
\\
\nonumber \\
\frac{\dd^2 \Iphi}{\dd \bart^2} &=& \left( - 3 \bH + \frac{\dd \ln{\Inm}}{\dd \bart} \right) \frac{\dd \Iphi}{\dd \bart} - \frac{1}{2 \Inm} \frac{\dd \Iv}{\dd \Iphi} ,
\label{EOM3A_JF}
\ea
where we have neglected the contributions of the matter part of the action since we assume that the energy density and pressure of the scalar field dominate during the inflationary epoch.

The standard HSRPs in the JF have the form~\cite{Kuusk2016}
\be 
\bar{\epsilon}_0 \equiv - \frac{1}{\bH^2} \frac{\dd \bH}{\dd \bart} = - \frac{\dd \ln{\bH}}{\dd \ln{\bar{a}}} , \qquad \bar{\eta} \equiv - \left( \bH \frac{\dd \Iphi}{\dd \bart} \right)^{-1} \frac{\dd^2 \Iphi}{\dd \bart^2}.
\label{eps_eta_JF}
\ee
Inflation in the JF occurs as long as $\bar{\epsilon}_0 < 1$, and slow rollover happens while $\bar{\epsilon}_0 \ll 1$. In the next section, we will be concerned with higher order corrections to the inflationary indices. As a result, we will need a series of slow-roll parameters which, following~\cite{Kuusk2016}, we take to be 
\be 
\bk_0 \equiv \frac{1}{\bH^2} \left( \frac{\dd \Iphi}{\dd \bart} \right)^2 = \left( \frac{\dd \Iphi}{\dd \ln{\bar{a}}} \right)^2 ,
\label{ka0_JF}
\ee
\be 
\bk_1 \equiv \frac{1}{\bH \bk_0} \frac{\dd \bk_0}{\dd \bart} = \frac{\dd \ln{\bk_0}}{\dd \ln{\bar{a}}} = 2 \left( - \bar{\eta} + \bar{\epsilon}_0 \right) ,
\label{ka1_JF}
\ee
\be 
\bk_{i+1} \equiv \frac{1}{\bH \bk_i} \frac{\dd \bk_i}{\dd \bart} = \frac{\dd \ln{\bk_i}}{\dd \ln{\bar{a}}}.
\label{kai+1_JF}
\ee
In the JF, it is also useful to consider a second series of slow-roll parameters involving the invariant $\Inm$ and thus related to the nonminimal coupling. This series has the form~\cite{Kuusk2016}
\be 
\bl_0 \equiv \frac{1}{2 \bH} \frac{\dd \ln{\Inm}}{\dd \bart} = \frac{1}{2} \frac{\dd \ln{\Inm}}{\dd \ln{\bar{a}}} ,
\label{la0_JF}
\ee
\be 
\bl_1 \equiv \frac{1}{\bH \bl_0} \frac{\dd \bl_0}{\dd \bart} = \frac{\dd \ln{\bl_0}}{\dd \ln{\bar{a}}} ,
\label{la1_JF}
\ee
\be 
\bl_{i+1} \equiv \frac{1}{\bH \bl_i} \frac{\dd \bl_i}{\dd \bart} = \frac{\dd \ln{\bl_i}}{\dd \ln{\bar{a}}} .
\label{lai+1_JF}
\ee
Now, using the definitions of the slow-roll parameters~\eqref{eps_eta_JF}-\eqref{lai+1_JF} we can rewrite the system of the field equations~\eqref{EOM1A_JF}-\eqref{EOM3A_JF} as
\ba 
\Iv &=& \bH^2 \Inm \left( 3 - \bk_0 - 6 \bl_0 + 3 \bl_0^2 \right) ,
\label{EOM1B_JF}
\\
\nonumber \\
\bk_0 &=& \beps_0 - \bl_0 \left( 1 + \beps_0 - \bl_0 - \bl_1 \right) ,
\label{EOM2B_JF}
\\ 
\nonumber \\
- \frac{1}{2\Inm} \frac{\dd \Iv}{\dd \Iphi} &=& \bH \frac{\dd \Iphi}{\dd \bart} \left( 3 - \beps_0 + \frac{1}{2} \bk_1 - 2 \bl_0 \right) .
\label{EOM3B_JF}
\ea
In the slow-roll regime we must have~\cite{Kuusk2016}
\be 
\vert \bk_0 \vert \ll 1 , \quad \vert \bk_1 \vert \ll 1 , \quad \vert \bl_0 \vert \ll 1 , \quad \vert \bl_1 \vert \ll 1 ,
\label{SR_cond_JF}
\ee
and then the slow-rolling inflaton obeys the following approximate equations:
\be 
\Iv \approx 3 \bH^2 \Inm , \qquad 3 \bH \frac{\dd \Iphi}{\dd \bart} \approx - \frac{1}{2 \Inm} \frac{\dd 
\Iv}{\dd \Iphi}.
\label{SR_EOMs_JF}
\ee

\subsection{Slow-roll in the Einstein frame}
\label{subsec:SREF}

Analogously to the JF, the field equations in terms of the invariants in the EF have the form~\cite{Kuusk2016}
\be 
\hH^2 = \frac{1}{3} \left[ \left( \frac{\dd \Iphi}{\dd \hatt} \right)^2 + \Iv \right] ,
\label{EOM1A_EF}
\ee
\be 
\frac{\dd \hH}{\dd \hatt} = - \left( \frac{\dd \Iphi}{\dd \hatt} \right)^2 ,
\label{EOM2A_EF}
\ee
\be 
\frac{\dd^2 \Iphi}{\dd \hatt^2} = - 3 \hH \frac{\dd \Iphi}{\dd \hatt} - \frac{1}{2} \frac{\dd \Iv}{\dd \Iphi} .
\label{EOM3A_EF}
\ee
The standard slow-roll parameters now are
\be
\heps_0 \equiv - \frac{1}{\hH^2} \frac{\dd \hH}{\dd \hatt} = - \frac{\dd \ln{\hH}}{\dd \ln{\hat{a}}} , \qquad  \hat{\eta} \equiv - \left( \hH \frac{\dd \Iphi}{\dd \hatt} \right)^{-1} \frac{\dd^2 \Iphi}{\dd \hatt^2},
\label{eps_eta_EF}
\ee
and again it will be useful to consider the following series of slow-roll parameters:
\be 
\hk_0 \equiv \frac{1}{\hH^2} \left( \frac{\dd \Iphi}{\dd \hatt} \right)^2 = \left( \frac{\dd \Iphi}{\dd \ln{\hat{a}}} \right)^2 ,
\label{ka0_EF}
\ee
\be 
\hk_1 \equiv \frac{1}{\hH \hk_0} \frac{\dd \hk_0}{\dd \hatt} = \frac{\dd \ln{\hk_0}}{\dd \ln{\hat{a}}} = 2 \left( - \hat{\eta} + \heps_0 \right) ,
\label{ka1_EF}
\ee
\be 
\hk_{i+1} \equiv \frac{1}{\hH \hk_i} \frac{\dd \hk_i}{\dd \hatt} = \frac{\dd \ln{\hk_i}}{\dd \ln{\hat{a}}} .
\label{kai+1_EF}
\ee
With the above definitions, the system~\eqref{EOM1A_EF}-\eqref{EOM3A_EF} can be rewritten as
\ba 
\Iv &=& \hH^2 \left( 3 - \hk_0 \right) ,
\label{EOM1B_EF}
\\
\nonumber \\
\hk_0 &=& \heps_0 ,
\label{EOM2B_EF}
\\
\nonumber \\ 
-\frac{1}{2} \frac{\dd \Iv}{\dd \Iphi} &=& \hH \frac{\dd \Iphi}{\dd \hatt} \left( 3 - \heps_0 + \frac{1}{2} \hk_1 \right) .
\label{EOM3B_EF}
\ea
The slow-roll conditions are now simply
\be 
\vert \hk_0 \vert \ll 1 , \qquad \vert \hk_1 \vert \ll 1 ,
\label{SR_cond_EF}
\ee
and the approximate forms of the equations~\eqref{EOM1B_EF}, \eqref{EOM3B_EF} become
\be 
\Iv \approx 3 \hH^2 , \qquad 3 \hH \frac{\dd \Iphi}{\dd \hatt} \approx - \frac{1}{2} \frac{\dd \Iv}{\dd \Iphi} .
\label{SR_EOMs_EF}
\ee

In the next section, we will calculate the inflationary indices up to third order in the slow-roll parameters in both the EF and JF and then compare the results. It will prove useful to relate the EF slow-roll parameters with the JF ones. This can be done using Eqs.~\eqref{trans_time_scale-factor}, \eqref{trans_Hubble}. We have
\be 
\hk_0 = \frac{\bk_0}{(1 - \bl_0)^2} , \qquad \hk_1 = \frac{\bk_1}{1 - \bl_0} + \frac{2 \bl_0 \bl_1}{(1 - \bl_0)^2} ,
\label{ka_EF-to-JF}
\ee
\be 
\heps_0 = \frac{\beps_0 - \bl_0}{1 - \bl_0} + \frac{\bl_0 \bl_1}{(1 - \bl_0)^2} .
\label{eps_JF-to-EF-to-JF}
\ee

\subsection{Invariant potential slow-roll parameters}
In the spirit of~\cite{Liddle1994}, we also define a hierarchy of slow-roll parameters in terms of the invariant inflaton potential. The standard potential slow-roll parameter $\epsilon_V$ assumes the form \cite{Kuusk2016}
\beq
\eps_V=\frac{1}{4 \Iv^2}\left(\frac{\dd \Iv}{\dd \Iphi} \right)^2 ,
\label{eps_V}
\eeq
while $\eta_V$ and higher-order parameters can be encoded in 
\beq
{}^n\beta_V \equiv \left( \frac{1}{2 \Iv} \right)^{n}\left( \frac{\dd \Iv}{\dd \Iphi} \right)^{n-1} \left( \frac{\dd^{(n+1)} \Iv}{\dd \Iphi^{(n+1)}} \right) ,
\label{hier_V}
\eeq 
where ${}^n\beta_V$ is a parameter of order $n$ in the slow-roll approximation. The first three parameters arising from this hierarchy are
\ba
\eta_V &=& \frac{1}{2 \Iv}\left( \frac{\dd^2 \Iv}{\dd \Iphi^2}\right) ,
\label{eta_V}
\\
\nonumber \\
\zeta_V^2 &=& \frac{1}{4 \Iv^2}\left(\frac{\dd \Iv}{\dd \Iphi} \right) \left(\frac{\dd^3 \Iv}{\dd \Iphi^3} \right) ,
\label{zeta_V}
\\
\nonumber \\
\rho_V^3 &=& \frac{1}{8 \Iv^3}\left( \frac{\dd^2 \Iv}{\dd \Iphi^2}\right) \left( \frac{\dd^4 \Iv}{\dd \Iphi^4}\right) .
\label{rho_V}
\ea
Note that we have changed the symbols $\xi$ and $\sigma$ of \cite{Liddle1994} in order to avoid confusion with the nonminimal coupling and one of the model functions, respectively.

\section{Higher-order spectral indices}
\label{sec:spectra-indices}

In this section, we compute the tensor and scalar power spectra up to second-order corrections in the slow-roll approximation and the corresponding spectral indices in both the JF and EF using the invariant slow-roll parameters of Secs~\ref{subsec:SRJF} and \ref{subsec:SREF}. We present the detailed calculation in the JF, and only give the final results for the EF since the calculation follows along the same lines with JF.

\subsection{Jordan frame analysis}
The evolution of linear (tensor and scalar) curvature cosmological perturbations in a flat FLRW background and in the presence of a scalar inflaton field is governed by the \textit{Mukhanov-Sasaki equation} (MSE) \cite{Sasaki1986, Mukhanov1988} which reads \cite{Hwang1990a, Hwang1996, Hwang1997a, Hwang1997, Hwang1998, Hwang1998a, Noh2001}
\be
\frac{\dd ^2 \nu}{\dd \tau ^2}+\left( k^2 -\frac{1}{z}\frac{\dd ^2 z}{\dd \tau^2}\right) \nu=0 ,
\label{Mukhanov-Sasaki-nu}
\ee
where $k$ corresponds to the scale of the Fourier mode $\mathbf{k}$ of the gauge-invariant \textit{comoving curvature perturbation} $\mathcal{R}_k$ \cite{Bardeen1980}. Furthermore, the field $\nu$ (usually referred to as the \textit{Mukhanov field}) is related to $\mathcal{R}_k$ via $\nu \equiv z \mathcal{R}_k$, where $z$ is a parametrization-independent quantity that depends on both the background and the type of perturbations~\cite{Kuusk2016}. For tensor perturbations,
\be 
z = \frac{\bar{a}}{\sqrt{\Inm}} = \hat{a} ,
\ee
while for scalar perturbations
\be 
z = \sqrt{\frac{2}{\Inm}} \frac{\bar{a}}{\bar{H} \left( 1 - \bl_0 \right)} \frac{\dd \Iphi}{\dd \bart} =  \sqrt{2} \frac{\hat{a}}{\hat{H}} \frac{\dd \Iphi}{\dd \hatt} .
\ee
Therefore, the evolution equation \eqref{Mukhanov-Sasaki-nu} is parametrization-independent and also has the same functional form for tensor and scalar perturbations. The two asymptotic solutions for the scalar field $\nu$ corresponding to the subhorizon and the superhorizon limit can be written respectively as
\beq
\nu\rightarrow \left\{
                \begin{array}{ll}
                  \frac{1}{\sqrt{2 k}} e^{-ik\tau} \;\; &\text{as} \;\; -k \tau \rightarrow \infty,\\
             A_k z \;\; &\text{as} \;\; -k \tau\rightarrow 0.
                \end{array}
              \right.
\label{MS-limits-nu}              
\eeq
The power spectrum for cosmological perturbations is usually defined by the two-point correlation function for $\mathcal{R}_k$ in the following way:
\beq
\left\langle \mathcal{R}_k,\mathcal{R}_{k'} \right\rangle  = (2 \pi)^2 \delta^3 (\mathbf{k}-\mathbf{k}') P_{\mathcal{R}}(k) ,
\eeq
where all quantities are calculated at the time when the mode $k$ crosses the horizon [when $k^{-1}$ equals the Hubble radius $(aH)^{-1}$]. Note that the horizon-crossing condition is not the same in the two frames. In the EF one has the condition $k=\hat{a}\hat{H}$ while in the JF using \eqref{trans_time_scale-factor},\eqref{trans_Hubble} and \eqref{la0_JF} one should use $k=\bar{a}\bar{H}(1-\bar{\lam}_0)$ to evaluate quantities at the time of horizon crossing. Now, using the relation between $\mathcal{R}_k$ and the Mukhanov field and the asymptotic superhorizon limit \eqref{MS-limits-nu} we can rewrite the power spectrum as
\beq
P(k) =\left( \frac{k^3}{2 \pi^2} \right) \lim_{-k \tau \rightarrow 0} \left| \frac{\nu}{z} \right|^2=\frac{k^3}{2 \pi^2} |A_k|^2.
\label{power-spectrum-def}
\eeq
This way the calculation of the spectrum reduces to simply finding the form of the amplitude of the field $\nu$ in the superhorizon limit. The MSE is usually solved in terms of Hankel functions by treating the slow-roll parameters as constant during inflation \cite{Stewart1993}. Since we want to obtain higher-order results for the power spectra and the spectral indices we cannot adhere to this assumption. Instead, we employ the Green's function method introduced by Stewart and Gong~\cite{Gong2001} which is valid to any order\footnote{See \cite{Stewart2002, Gong2004, Kim2004, Wei2004, Kim2005, Kadota2005, Joy2005, Dvorkin2011, Adshead2011, Kumazaki2011, Miranda2012, Adshead2013, BeltranJimenez2013, Adshead2014, Gong2014, Achucarro2014, Motohashi2015, Motohashi2017} for various extensions and applications of this method and \cite{Schwarz2001, Martin2003, Casadio2005, Casadio2005a, Casadio2005b, Habib2004, Habib2005, Lorenz2008, Zhu2014, Zhu2014b, Alinea2016, Rojas2009, Rojas2012} for other related methods.}. 

Now, in order to compute $A_k$ one has to solve the MSE \eqref{Mukhanov-Sasaki-nu} which is a second-order differential equation. Thus in order to uniquely specify the solution for the field $\nu$ the use of two boundary conditions is necessary. To this end, one can use the asymptotic solutions \eqref{MS-limits-nu} as boundary conditions. By introducing the dimensionless variable $x \equiv -k \tau$ and redefining the field as $y \equiv \sqrt{2 k} \nu$, the asymptotic solutions become
\beq
y\rightarrow \left\{
                \begin{array}{ll}
                   e^{-ix} \;\; &\text{as} \;\; x \rightarrow \infty,\\
              \sqrt{2 k} A_k z \;\; &\text{as} \;\; x \rightarrow 0.
                \end{array}.
              \right.
\label{MS-limits-y}              
\eeq
Also, by assuming the following ansatz for $z$:
\beq
z=\frac{1}{x}f(\ln{x}) ,
\label{ansatz}
\eeq
we can recast the MSE in the form
\beq
\frac{\dd ^2y}{\dd x^2}+\left(1-\frac{2}{x^2} \right) y=\frac{1}{x^2} g(\ln{x})y ,
\label{Mukhanov-Sasaki-y}
\eeq
where the function $g$ is defined through
\beq
g(\ln{x})=\frac{1}{f(\ln{x})} \left[-3 \frac{\dd f(\ln{x})}{\dd \ln{x}}+ \frac{\dd ^2 f(\ln{x})}{\dd (\ln{x})^2} \right].
\label{g-func-def}
\eeq
The homogeneous solution with the appropriate asymptotic behavior at $x \rightarrow \infty$ is
\beq
y_0(x)=\left(1+\frac{i}{x} \right) e^{i x}.
\label{hom-sol}
\eeq
By ``appropriate behavior" we mean that \eqref{hom-sol} reduces to the usual Minkowski modes in the deep subhorizon regime.
Combining \eqref{Mukhanov-Sasaki-y} and \eqref{MS-limits-y} we can rewrite the MSE as an integral equation
\beq
y(x)=y_0(x)+\frac{i}{2}\int_x^{\infty} \dd u \frac{1}{u^2} g (\ln{u})y(u) \left[ y_0^*(u) y_0(x)-y_0^*(x) y_0(u) \right]
\label{int-sol}
\eeq
and seek a perturbative solution to \eqref{int-sol}. We start by Taylor-expanding $xz$ around $x=1$ in the following way: 
\beq
xz=f(\ln{x})=\sum_{n=0}^{\infty} \frac{f_n}{n!}(\ln{x})^n,
\label{xz-expansion}
\eeq
where the $n$--th order coefficient of the expansion is of the same order in slow-roll and is given by
\beq
f_n=\frac{\dd ^n (xz)}{\dd (\ln{x})^n}.
\label{f-coeff}
\eeq
In terms of the slow-roll parameters 
\beq
\bar{\epsilon}_n=\frac{(-1)^{n+1}}{\bar{H}} \frac{\bar{H}^{(n+1)}}{\bar{H}^{(n)}}
\label{eps-hier}
\eeq
we can expand the conformal time up to second order corrections and thus have the following approximation~\cite{Alinea2016}:
\beq
x=-k\tau = -k\int\frac{\dd \bar{t}}{\bar{a}} = \frac{k}{\bar{a}\bar{H}}(1+\bar{\eps}_0+3 \bar{\eps}_0^2+\bar{\eps}_0 \bar{\eps}_1) .
\label{x-expansion-eps}
\eeq
Then, using the relations
\beq
\bar{\eps}_0=\bar{\lam}_0+\frac{\bar{\kap}_0}{(1-\bar{\lam}_0)}-\frac{\bar{\lam}_0 \bar{\lam}_1}{(1-\bar{\lam}_0)} ,
\label{eps-to-kappa1}
\eeq
\beq
\bar{\eps}_0^2 \approx \bar{\lam}_0^2+\frac{\bar{\kap}_0^2}{(1-\bar{\lam}_0)^2}+2\frac{\bar{\lam}_0\bar{\kap}_0}{(1-\bar{\lam}_0)} ,
\label{eps-to-kappa2}
\eeq
\beq
2 \bar{\eps}_0^2+\bar{\eps}_0 \bar{\eps}_1 \approx \bar{\lam}_0 \bar{\lam}_1+\frac{\bar{\kap}_0\bar{\kap}_1}{(1-\bar{\lam}_0)} ,
\label{eps-to-kappa3}
\eeq
we can express $x$ in terms of the $\bk$ and $\bl$ slow-roll parameters,
\beq
x = \frac{k}{\bar{a}\bar{H}} \left( 1+\bar{\lam}_0+\bar{\kap}_0+3\bar{\lam}_0\bar{\kap}_0+\bar{\kap}_0\bar{\kap}_1+\bar{\kap}_0^2+\bar{\lam}_0^2\right) .
\label{x-expansion-kappa}
\eeq
The second-order power spectrum is then given in terms of the coefficients $f_0$, $f_1$ and $f_2$ as~\cite{Gong2001}
\be
P(k)=\frac{k^2}{(2\pi)^2}\frac{1}{f_0^2}\left[1-2 \alp \, \frac{f_1}{f_0}+\left( 3 \alp^2 -4 +\frac{5 \pi^2}{12}\right) \left( \frac{f_1}{f_0}\right)^2 +\left( -\alp^2+\frac{\pi^2}{12}\right) \frac{f_2}{f_0}\right],
\label{power-spectrum-result}
\ee
where $\alp  \equiv (2-\ln{2}-\gamma) \simeq 0.729637$ and $\gamma \simeq 0.577216$ is the Euler-–Mascheroni constant~\cite{Stewart2002}.
For tensor perturbations in the JF we have that up to second order terms 
\ba
f^{T}_0 &=& \left.\frac{k}{\bar{H} \sqrt{\Inm}}\left(  1+\bar{\lam}_0+\bar{\kap}_0+3\bar{\lam}_0\bar{\kap}_0+\bar{\kap}_0\bar{\kap}_1+2\bar{\kap}_0^2+\bar{\lam}_0^2\right)\right\vert_{k= \bar{a} \bar{H} (1-\bar{\lam}_0)} ,
\label{p0-coeff}
\\
\nonumber \\
f^{T}_1 &=& \left.\frac{k}{\bar{H} \sqrt{\Inm}} \left(  -\bar{\kap}_0-3\bar{\kap}_0\bar{\lam}_0-2\bar{\kap}_0^2-\bar{\kap}_0\bar{\kap}_1 \right)\right\vert_{k= \bar{a} \bar{H} (1-\bar{\lam}_0)} ,
\label{p1-coeff}
\\
\nonumber \\
f^{T}_2 &=& \left.\frac{k}{\bar{H} \sqrt{\Inm}} \left(  \bar{\kap}_0^2+\bar{\kap}_0\bar{\kap}_1 \right)\right\vert_{k= \bar{a} \bar{H} (1-\bar{\lam}_0)},
\label{p2-coeff}
\ea
where the slow-roll parameters are evaluated at the time of the horizon crossing. We have also introduced the superscript ``T" to discriminate from the corresponding coefficients of the scalar perturbations which will be denoted by an ``S".

Substitution of these coefficients into \eqref{power-spectrum-result} results in the following expression for the second order corrected tensor power spectrum in the slow-roll approximation:
\be 
\begin{split}
\bar{P}_{T} =
\left[ \frac{\bar{H}^2\Inm}{(2\pi)^2} \right] & \left[  1-2\bar{\lam}_0+(2\alp-2)\bar{\kap}_0+\bar{\lam}_0^2+\left( 2 \alp^2-2\alp-5+\frac{\pi^2}{2} \right)\bar{\kap}_0^2
\right. \\
  &\quad \left. {}
 +\left( -\alp^2+2\alp-2+\frac{\pi^2}{12}\right)\bar{\kap}_0\bar{\kap}_1 \right].
\end{split}
\label{tensor-spectrum-JF}
\ee
The tensor spectral index is defined as the logarithmic derivative of the power spectrum
\beq
\bar{n}_{T} \equiv \frac{\dd \ln{\bar{P}_{T}(k)}}{\dd \ln{k}}
\eeq
and thus the third order JF tensor scalar spectral index is obtained to be
\be 
\begin{split}
\bar{n}_{T} =&-2\bar{\kap}_0-2\bar{\kap}_0^2-4\bar{\lam}_0 \bar{\kap}_0+(2 \alp-2)\bar{\kap}_0\bar{\kap}_1-6\bar{\lam}_0^2\bar{\kap}_0+(4 \alp-2)\bar{\lam}_0\bar{\lam}_1\bar{\kap}_0-8\bar{\lam}_0\bar{\kap}_0^2
\\
&  +(6 \alp-6)\bar{\lam}_0\bar{\kap}_0\bar{\kap}_1-2\bar{\kap}_0^3+(6 \alp-16+\pi^2)\bar{\kap}_0^2\bar{\kap}_1+\left(-\alp^2+ 2 \alp-2+\frac{\pi^2}{12} \right) (\bar{\kap}_0\bar{\kap}_1^2+\bar{\kap}_0\bar{\kap}_1\bar{\kap}_2).
\end{split}
\label{tensor-index-JF}
\ee
For scalar perturbations in the JF the coefficients $f^{S}$ are slightly more complicated than their $f^{T}$ counterparts and have the following second order forms:
\ba
f^{S}_0 &=& \frac{k}{\bar{H}^2}\sqrt{\frac{2}{\Inm}}\frac{\dd  \Iphi }{\dd \bar{t}}\left.\left[ 
1+2\bar{\lam}_0+\bar{\kap}_0+4\bar{\lam}_0\bar{\kap}_0+\frac{3}{2}\bar{\kap}_0\bar{\kap}_1+2\bar{\kap}_0^2+3\bar{\lam}_0^2 \right]\right\vert_{k= \bar{a} \bar{H} (1-\bar{\lam}_0)} ,
\label{f0-coeff}
\\
\nonumber \\
f^{S}_1 &=& -\frac{k}{\bar{H}^2}\sqrt{\frac{2}{\Inm}}\frac{\dd  \Iphi }{\dd \bar{t}} \left. \left[ \bar{\kap}_0+\frac{\bar{\kap}_1}{2}+2\bar{\kap}_0\bar{\kap}_1+4\bar{\kap}_0\bar{\lam}_0+\frac{3}{2}\bar{\lam}_0\bar{\kap}_1+\bar{\lam}_0\bar{\lam}_1+2\bar{\kap}_0^2 \right]\right\vert_{k= \bar{a} \bar{H} (1-\bar{\lam}_0)} ,
\label{f1-coeff}
\\
\nonumber \\
 f^{S}_2 &=& \frac{k}{\bar{H}^2}\sqrt{\frac{2}{\Inm}}\frac{\dd  \Iphi }{\dd \bar{t}}\left.\left[ \frac{\bar{\kap}_1^2}{4}+2\bar{\kap}_0\bar{\kap}_1+\bar{\kap}_0^2+\frac{\bar{\kap}_1\bar{\kap}_2}{2} \right]\right\vert_{k= \bar{a} \bar{H} (1-\bar{\lam}_0)} .
\label{f2-coeff}
\ea
Then the scalar power spectrum in the JF is
\be 
\begin{split}
\bar{P}_{S} =
\left[ \frac{\bar{H}^4}{(2\pi)^2}\frac{\Inm}{2} \left(\frac{\dd  \Iphi }{\dd\bar{t}}\right)^{-2}  \right]& \left[  1-4\bar{\lam}_0+(2\alp-2)\bar{\kap}_0+\alp\bar{\kap}_1+\left(2\alp^2-2 \alp-5+\frac{\pi^2}{2} \right)\bar{\kap}_0^2
\right. \\
  &\quad \left. {}
+(4-4\alp)\bar{\lam}_0\bar{\kap}_0+(-3\alp)\bar{\lam}_0\bar{\kap}_1+\left( \frac{\alp^2}{2}-1+\frac{\pi^2}{8}\right)\bar{\kap}_1^2+6\bar{\lam}_0^2
\right. \\
  &\quad \left. {}
+2\bar{\alp}\bar{\lam}_0\bar{\lam}_1+\left( \alp^2+\alp-7+\frac{7\pi^2}{12} \right)\bar{\kap}_0\bar{\kap}_1+ \left( -\frac{\alp^2}{2}+\frac{\pi^2}{24} \right)\bar{\kap}_1\bar{\kap}_2 \right].
\label{scalar-spectrum-JF}
\end{split}
\ee
Substitution of the latter in the definition of the scalar spectral index
\beq
\bar{n}_{S} \equiv1+ \frac{\dd \ln{\bar{P}_{S}}}{\dd \ln{k}}
\eeq
results in the following third order expression for the scalar index in the JF:
\be 
\begin{split}
\bar{n}_{S} =&
1-2 \bar{\kap}_0-\bar{\kap}_1-2\bar{\kap}_0^2-2\bar{\lam}_0\bar{\lam}_1+\alp\bar{\kap}_1\bar{\kap}_2-\bar{\kap}_1\bar{\lam}_0-4\bar{\kap}_0\bar{\lam}_0+(2\alp-3)\bar{\kap}_1 \bar{\kap}_0-2\bar{\kap}_0^3-8\bar{\lam}_0\bar{\kap}_0^2
 \\  
& -6\bar{\lam}_0^2\bar{\kap}_0+(6\alp-17+\pi^2)\bar{\kap}_0^2\bar{\kap}_1-\bar{\kap}_1\bar{\lam}_0^2+\left( -2+\frac{\pi^2}{4} \right)\bar{\kap}_1^2\bar{\kap}_2-4\bar{\lam}_0^2\bar{\lam}_1+2\alp\bar{\lam}_0 \bar{\lam}_1^2
\\
& +\left(-\frac{\alp^2}{2}+ \frac{\pi^2}{24} \right)\bar{\kap}_1\bar{\kap}_2^2+\left( -\alp^2+3 \alp-7+\frac{7 \pi^2}{12} \right)\bar{\kap}_0\bar{\kap}_1^2 +2\alp\bar{\lam}_0\bar{\lam}_1\bar{\lam}_2+(6\alp-9)\bar{\lam}_0\bar{\kap}_0\bar{\kap}_1
\\
& +(4\alp-4)\bar{\lam}_0\bar{\lam}_1\bar{\kap}_0+(\alp+1)\bar{\kap}_1\bar{\lam}_0\bar{\lam}_1+2\alp\bar{\lam}_0\bar{\kap}_1\bar{\kap}_2
+ \left( -\frac{\alp^2}{2}+\frac{\pi^2}{24} \right)\bar{\kap}_1\bar{\kap}_2\bar{\kap}_3
\\
& +\left(-\alp^2+ 4\alp-7+\frac{7 \pi^2}{12} \right)\bar{\kap}_0\bar{\kap}_1\bar{\kap}_2 .
\end{split}
\label{scalar-index-JF}
\ee
Finally, with the higher order corrected expressions for the power spectra for scalar and tensor perturbations in the JF at our disposal, it is trivial to compute the tensor-to-scalar ratio,
\be
\begin{split}
\bar{r} = & 16 \bar{\kap}_0 \left[ 1+2\bar{\lam}_0-\alp\bar{\kap}_1+3\bar{\lam}_0^2-2\alp \bar{\lam}_0\bar{\lam}_1-3\alp\bar{\lam}_0\bar{\kap}_1 +\left(-\alp+5-\frac{\pi^2}{2} \right)\bar{\kap}_0\bar{\kap}_1
\right. \\
  & \qquad \left. {}
+\left( \frac{\alp^2}{2}+1-\frac{\pi^2}{8}\right)\bar{\kap}_1^2+\left( \frac{\alp^2}{2}-\frac{\pi^2}{24} \right)\bar{\kap}_1\bar{\kap}_2  \right] .
\end{split}
\label{ratio-JF}
\ee

\subsection{Einstein frame results}
Repeating the same analysis in the EF, we obtain the tensor power spectrum
\be
\hat{P}_{T}=\frac{\hat{H}^2}{(2 \pi)^2} \left[ 1+(2\alp-2)\hat{\kap}_0+\left( 2\alp^2-2\alp-5+\frac{\pi^2}{2}\right)\hat{\kap}_0^2+\left(-\alp^2+ 2\alp-2+\frac{\pi^2}{12} \right)\hat{\kap}_0\hat{\kap}_1 \right] ,
\label{tensor-spectrum-EF}
\ee
the tensor spectral index
\be 
\begin{split}
\hat{n}_{T} =&-2 \hat{\kap}_0-2\hat{\kap}_0^2+(2 \alp-2) \hat{\kap}_0 \hat{\kap}_1-2 \hat{\kap}_0^3+(6 \alp-16+\pi^2)\hat{\kap}_0^2 \hat{\kap}_1
 \\  
& +\left(-\alp^2+ 2 \alp-2+\frac{\pi^2}{12} \right) (\hat{\kap}_0 \hat{\kap}_1^2+\hat{\kap}_0\hat{\kap}_1\hat{\kap}_2) ,
\end{split}
\label{tensor-index-EF}
\ee
the scalar power spectrum
\be 
\begin{split}
\hat{P}_{S} =
\left[ \frac{\hat{H}^4}{2(2\pi)^2} \left(\frac{\dd  \Iphi }{\dd \hat{t}}\right)^{-2}  \right]& \left[  1+(2\alp-2)\hat{\kap}_0+\alp
\hat{\kap}_1+\left( 2\alp^2-2\alp-5+\frac{\pi^2}{2} \right)\hat{\kap}_0^2
\right. \\
  &\quad \left. {}
+\left( \frac{\alp^2}{2}-1+\frac{\pi^2}{8}\right)\hat{\kap}_1^2+\left( \alp^2+\alp-7+\frac{7 \pi^2}{12} \right)\hat{\kap}_0\hat{\kap}_1
\right. \\
  &\quad \left. {}
+\left( -\frac{\alp^2}{2}+\frac{\pi^2}{24} \right)\hat{\kap}_1\hat{\kap}_2 \right] ,
\label{scalar-spectrum-EF}
\end{split}
\ee
the scalar spectral index
\be 
\begin{split}
\hat{n}_{S} =&
1-2 \hat{\kap}_0-\hat{\kap}_1-2\hat{\kap}_0^2+\alp\hat{\kap}_1\hat{\kap}_2+(2\alp-3)\hat{\kap}_0 \hat{\kap}_1-2\hat{\kap}_0^3+(6\alp-17+\pi^2)\hat{\kap}_0^2\hat{\kap}_1 
\\
& +\left(-2+ \frac{\pi^2}{4} \right)\hat{\kap}_1^2\hat{\kap}_2+\left( -\frac{\alp^2}{2} +\frac{\pi^2}{24}\right)\hat{\kap}_1\hat{\kap}_2^2+\left( -\alp^2+3 \alp-7+\frac{7 \pi^2}{12} \right)\hat{\kap}_0\hat{\kap}_1^2
\\
& 
+ \left(-\frac{\alp^2}{2}+ \frac{\pi^2}{24} \right)\hat{\kap}_1\hat{\kap}_2\hat{\kap}_3+\left( -\alp^2+4\alp-7+\frac{7 \pi^2}{12} \right)\hat{\kap}_0\hat{\kap}_1\hat{\kap}_2 ,
\end{split}
\label{scalar-index-EF}
\ee
and finally the tensor-to-scalar ratio
\be
\hat{r} = 16 \hat{\kap}_0 \left[ 1-\alp\hat{\kap}_1 +\left(-\alp+5-\frac{\pi^2}{2} \right)\hat{\kap}_0\hat{\kap}_1+\left( \frac{\alp^2}{2}+1-\frac{\pi^2}{8}\right)\hat{\kap}_1^2
+\left( \frac{\alp^2}{2}-\frac{\pi^2}{24} \right)\hat{\kap}_1\hat{\kap}_2  \right] .
\label{ratio-EF}
\ee
Note that the above results have been obtained using the condition $k = \hat{a} \hat{H}$ at the time of horizon crossing.

\subsection{Equivalence of the frames up to third order}

It has been reported by the authors of \cite{Kuusk2016} that the EF and JF spectral indices are equivalent up to second order in the slow-roll expansion. In this work we have obtained the third-order corrected expressions for the indices in the two frames. It is thus intriguing to see whether this equivalence extends to the third-order expressions also. Expanding the EF slow-roll parameters \eqref{ka_EF-to-JF} up to third order in the JF slow-roll parameters we have 
\be 
\hk_0 \approx \bk_0 + 2 \bk_0 \bl_0 + 3 \bk_0 \bl_0^2 ,
\label{hk0-to-bk0}
\ee
\be 
\hk_1 \approx \bk_1 + \bk_1 \bl_0 + \bk_1 \bl_0^2 + 2 \bl_0 \bl_1 + 4 \bl_0^2 \bl_1 ,
\label{hk1-to-bk1}
\ee
\be 
\hk_1 \hk_2 \approx \bk_1 \bk_2 + 2 \bk_1 \bk_2 \bl_0 + \bk_1 \bl_0 \bl_1 + 2 \bl_0 \bl_1^2 + 2 \bl_0 \bl_1 \bl_2 ,
\label{hk1hk2-to-bk1bk2}
\ee
\be 
\hk_0 \hk_1 \hk_2 \approx \bk_0 \bk_1 \bk_2 \, , \quad  \hk_1 \hk_2 \hk_3 \approx \bk_1 \bk_2 \bk_3 .
\label{tripleh-to-tripleb}
\ee
Then, plugging \eqref{hk0-to-bk0} - \eqref{tripleh-to-tripleb} in the EF expressions for the indices \eqref{tensor-index-EF} - \eqref{ratio-EF} we find
\be 
\hat{n}_T = \bar{n}_T ,
\ee
\be 
\hat{n}_S = \bar{n}_S ,
\ee
\be 
\hat{r} = \bar{r} .
\ee
Therefore, the spectral indices calculated in the EF and JF coincide. Finally, since the Green's function method is valid up to arbitrary order in the slow-roll expansion, we expect the equivalence between the spectral indices in the JF and EF to also hold to all orders.

\subsection{Invariant expressions for the inflationary observables}
So far we have obtained the spectral indices and the tensor-to-scalar ratio in both the EF and JF. We have also shown that up to third order in the slow-roll expansion the results in the two frames are equivalent. We can take advantage of this equivalence and write down expressions for the inflationary observables only in terms of the invariant potential and its derivatives. The equivalence between the two frames allows then one to rewrite the EF results in terms of the invariant PSRPs and expect these results to hold in the JF too. In order to express the spectral indices in terms of the PSRPs defined in \eqref{eps_V} - \eqref{rho_V} we first use the following relations between the EF HSRPs \eqref{ka0_EF} - \eqref{kai+1_EF} and the ones defined in~\cite{Liddle1994}:
\begin{eqnarray}
\hat{\kap}_0 &=& \eps_H ,\\
\hat{\kap}_1 &=& -2 \eta_H+2 \eps_H ,\\
\hat{\kap}_1\hat{\kap}_2 &=& 4 \eps_H^2-6\eps_H \eta_H+2\zeta_H^2 ,\\
\hat{\kap}_1 \hat{\kap}_2^2+\hat{\kap}_1\hat{\kap}_2\hat{\kap}_3 &=& 16 \eps_H^3-22\eps_H^2 \eta_H+12\eps_H \eta_H^2+10 \eps_H \zeta_H^2-2 \eta_H \zeta_H^2-2 \rho_H^3.
\end{eqnarray}
Then, using the third-order Taylor expansions of the HSRPs in terms of the PSRPs~\cite{Liddle1994}, presented in Appendix \ref{app1:HSRPs-to-PSRPs}, we obtain the inflationary indices up to third order in the PSRPs
\be 
\begin{split}
n_{T} =&
 -2 \eps_V +\left( 8 \alp-\frac{22}{3} \right) \eps_V^2- \left( 4 \alp-\frac{8}{3} \right) \eps_V \eta_V+ \left(-32 \alp^2+\frac{189}{3}\alp-\frac{996}{9}+\frac{20 \pi^2}{3} \right) \eps_V^3
\\
& +\left( -4 \alp^2+4 \alp-\frac{46}{9}+\frac{\pi^2}{3}\right) \eps_V \eta^2_V +\left( 28 \alp^2-44 \alp +68 -\frac{13 \pi^2}{3}  \right) \eps_V^2 \eta_V
\\
& 
+ \left( -2 \alp^2+\frac{8}{3} \alp-\frac{28}{9}+\frac{\pi^2}{6}\right)\eps_V \zeta_V^2 ,
\end{split}
\label{tensor-index-V}
\ee
\be 
\begin{split}
n_{S} =&1
 -6 \eps_V+2\eta_V+\left(24 \alp -\frac{10}{3} \right)\eps^2_V-\left( 16 \alp+2\right)\eps_V \eta_V+\frac{2}{3} \eta_V^2+\left( 2 \alp+\frac{2}{3}\right) \zeta_V^2
\\
& -\left(90 \alp^2-\frac{104}{3}\alp+\frac{3734}{9}-\frac{87 \pi^2}{2} \right)\eps_V^3+\left( 90 \alp^2+\frac{4}{3} \alp+\frac{1190}{3}-\frac{87\pi^2}{2} \right) \eps_V^2 \eta_V
\\
& 
- \left( 16 \alp^2+12\alp+\frac{742}{9}-\frac{28 \pi^2}{3}\right) \eps_V \eta_V^2-\left(12 \alp^2+4 \alp+\frac{98}{3}-4 \pi^2 \right)\eps_V \zeta_V^2
\\
& 
+\left(\alp^2+\frac{8}{3}\alp+\frac{28}{3}-\frac{13 \pi^2}{2} \right) \eta_V \zeta^2_V+\frac{4}{9}\eta_V^3+\left(\alp^2+\frac{2}{3}\alp+\frac{2}{9}-\frac{\pi^2}{12} \right) \rho_V^3 ,
\end{split}
\label{scalar-index-V}
\ee
\be 
\begin{split}
r =&
16 \eps_V \left[1-\left(4 \alp+\frac{4}{3} \right) \eps_V+\left( 2\alp +\frac{2}{3}\right) \eta_V+\left(16 \alp^2+\frac{28}{3}\alp+\frac{356}{9}-\frac{14 \pi^2}{3} \right) \eps_V^2
\right. \\
  &\quad \left. {}
-\left(14 \alp^2+10 \alp+\frac{88}{3}-\frac{7\pi^2}{2} \right)\eps_V \eta_V+\left(2 \alp^2+2 \alp+\frac{41}{9}-\frac{\pi^2}{2} \right) \eta_V^2
\right. \\
  &\quad \left. {}
+\left(\alp^2+\frac{2}{3}\alp+\frac{2}{9}-\frac{\pi^2}{12} \right)\zeta_V^2 \right]
\end{split}
\label{ratio-V}
\ee

In a given model, once we derive the invariant potential $\Iv$ in terms of the invariant $\Iphi$, we can readily obtain the PSRPs and express the inflationary observables in an invariant way in terms of $\Iv$ and its derivatives.

\section{Number of $e$-folds}
\label{sec:efolds}

In this section, we consider the difference between the definitions for the number of $e$-folds in the EF and JF and study how it affects the values of the observables. Furthermore, we discuss various approaches for a more accurate determination of the value of the inflaton field at the end of inflation.

\subsection{Einstein vs Jordan}

The number of $e$-folds is usually defined in the EF as
\be 
\dd \hat{N} \equiv \hH \dd \hatt = \dd \ln \hat{a} = - \frac{1}{\sqrt{\hk_0}} \, \dd \Iphi = - \frac{1}{\sqrt{\heps_0}} \, \dd \Iphi = - \frac{1}{\sqrt{\epsilon_H}} \, \dd \Iphi .
\label{Eefolds}
\ee
Using \eqref{trans_time_scale-factor} the number of $e$-folds in the JF becomes
\be 
\dd \bar{N} = \dd \hat{N} + \frac{1}{2} \, \dd \ln \Inm = \left( - \frac{1}{\sqrt{\epsilon_H}} + \frac{1}{2} \frac{\dd \ln \Inm}{\dd \Iphi} \right) \dd \Iphi .
\label{Jefolds}
\ee
We see that the definitions for the number of $e$-folds in the two frames differ by the invariant factor $\frac{1}{2} \, \dd \ln \Inm$ which includes the nonminimal coupling in a given theory. Of course, when the scalar field is minimally coupled to gravity the two definitions coincide. Therefore, in general, the same number of $e$-folds in the two frames will translate to different values for the invariant $\Iphi$. This means that we will get different predictions for the observables depending on whether we use \eqref{Eefolds} or \eqref{Jefolds}. Typically the difference is small, but still comparable to (if not larger than) the difference for the observables if one chooses to use the first, second or third order results for $n_S$ and $r$ in terms of the slow-roll parameters. Furthermore, these types of differences can play a significant role in the future, with the advent of more precise measurements \cite{Matsumura2013, Finelli2016}, in regards to the characterization of an inflationary model as viable or not.

In order to quantify the aforementioned effects, we will next consider the nonminimal Coleman-Weinberg model introduced in \cite{Kannike2016}. The model functions are
\ba 
\mathcal{A} (\Phi) &=& \xi \Phi^2 ,
\label{CW-A}
\\
\mathcal{B} (\Phi) &=& 1 ,
\label{CW-B}
\\
\sigma(\Phi) &=& 0 ,
\label{CW-sigma}
\\
\mathcal{V} (\Phi) &=& \Lambda^4 + \frac{1}{8} \beta_{\lambda_\Phi} \left( \ln \frac{\Phi^2}{v^2_\Phi} - \frac{1}{2} \right) \Phi^4 ,
\label{CW-V}
\ea
where the cosmological constant $\Lambda^4$ was included in order to realize $\mathcal{V} (v_\Phi) = 0$ and $\beta_{\lambda_\Phi}$ is the beta function of the quartic scalar coupling $\lambda_\Phi$. Furthermore, in this model the Planck scale is dynamically generated through the VEV of the scalar field $v_\Phi$ and we have
\be 
1= \xi v^2_\Phi .
\label{VEV}
\ee
Minimization of the potential \eqref{CW-V} yields
\be 
\beta_{\lambda_\Phi} = 16 \, \frac{\Lambda^4}{v^4_\Phi} . 
\ee 
This means we can eliminate $\beta_{\lambda_\Phi}$ in \eqref{CW-V} and rewrite the potential as 
\be 
\mathcal{V} (\Phi) = \Lambda^4 \left\lbrace 1 + \left[ 2 \ln \left( \frac{\Phi^2}{v^2_\Phi} \right) - 1 \right] \frac{\Phi^4}{v^4_\Phi}   \right\rbrace 
\ee
From the expressions of the model functions \eqref{CW-A} - \eqref{CW-V} we can readily obtain the invariants $\Inm$, $\Iv$ and $\Iphi$. The invariant field takes the form
\be 
\Iphi = \sqrt{\frac{1 + 6 \xi}{2 \xi}} \ln \left( \frac{\Phi}{v_\Phi} \right) .
\ee
By inverting the above equation we can express the invariant $\Inm$ in terms of $\Iphi$ as
\be 
\Inm = e^{- 2 \sqrt{\frac{2 \xi}{1 + 6 \xi}} \Iphi } ,
\ee
and also the invariant potential $\Iv$ in terms of $\Iphi$ as
\be  
\Iv = \Lambda^4 \left( 4 \sqrt{\frac{2 \xi}{1 + 6 \xi}} \, \Iphi +  e^{- 4 \sqrt{\frac{2 \xi}{1 + 6 \xi}} \Iphi }  - 1 \right) ,
\label{CW-Iv}
\ee
where we used \eqref{VEV}. From the invariant potential \eqref{CW-Iv} we can calculate the PSRPs \eqref{eps_V}, \eqref{eta_V} - \eqref{rho_V} and then the scalar index $n_S$ [c.f. \eqref{scalar-index-V}] and the tensor-to-scalar ratio $r$ [c.f. \eqref{ratio-V}] and compare them with the experimental bounds. Another important observable is the amplitude of scalar perturbations $A_S = \left( 2.14 \pm 0.05 \right) \times 10^{-9}$ \cite{Ade2016a}, which can be used to constrain the value of $\Lambda$ (see Fig. 3 in \cite{Kannike2016}).

Now, depending on whether the field $\Phi$ rolls down from values larger or smaller than its VEV, the invariant $\Iphi$ can have positive or negative values. Since negative field inflation produces $r \gtrsim 0.15$ \cite{Kannike2016}, which is excluded by observations \cite{Ade2015, Ade2016}, we will not consider it further. Instead, we will only focus on positive field inflation which interpolates between quadratic \cite{Linde1983a} and linear \cite{McAllister2010} inflation depending on the value of the nonminimal coupling $\xi$. In the limit $\xi \rightarrow 0$, the invariant potential is approximated as
\be 
\Iv \vert_{\xi \rightarrow 0} \sim 16 \, \xi \, \Lambda^4 \, \Iphi^2 ,
\ee
while in the limit $\xi \rightarrow \infty$,
\be 
\Iv \rvert_{\xi \rightarrow \infty} \sim \frac{4}{\sqrt{3}} \, \Lambda^4 \, \Iphi .
\ee

Quadratic inflation is excluded by the Planck and BICEP2/Keck results \cite{Ade2015, Ade2016} but linear inflation  still lies within the $2 \sigma$ allowed region. In Table \ref{table:E-vs-J-efolds} we present our results for the first and third order scalar index $n_S$ and tensor-to-scalar ratio $r$ for various values of the nonminimal coupling $\xi$. For simplicity, we have assumed that inflation ends at $\Phi = v_\Phi$, or equivalently $\Iphi^{\rm end} = 0$, where the two frames coincide. Furthermore, we have approximated $\epsilon_H \approx \epsilon_V$ in the expressions \eqref{Eefolds} and \eqref{Jefolds}. In each case, for every value of $\xi$ considered, we have varied $\Iphi^{\rm HC}$ at horizon crossing in order to get $\hat{N} = 60$ and $\bar{N} = 60$. This means that we obtain a different value for $\Iphi$ depending on which definition for the $e$-folds we use. Consequently, the predictions for $n_S$ and $r$ differ. For small $\xi$ the difference between the frames is negligible. However, for larger $\xi$ the difference grows and becomes around $0.002$ (or $0.2 \%$) for $n_S$ and $0.005$ (or $8 \%$) for $r$ around $\xi = 10$. For large $\xi$, such a difference is actually larger than the difference between the first and third order results for the observables ($0.03 \%$ for $n_S$ and $1.9 \%$ for $r$). Both of these types of differences however should be within the reach of future experiments such as CORE and LiteBIRD \cite{Matsumura2013, Finelli2016} which are expected to measure $r$ with an accuracy of $10^{-3}$.

\begin{table}
\begin{center}
\begin{tabular}{| c | c | c | c | c | c |}
\hline
 & & & & & \\[-1em]
 & $n_S^{(\rm I)}$ & $n_S^{(\rm III)}$ & $r^{(\rm I)}$ & $r^{(\rm III)}$ & $\xi$ \\
\hline 
 & & & & & \\[-1em] 
$\hat{N} = 60$ & 0.96702 & 0.96712 & 0.12782 & 0.12552 & $10^{-5}$ \\
\hline
 & & & & & \\[-1em]
$\bar{N} = 60$ & 0.96699 & 0.96709 & 0.12792 & 0.12562 & $10^{-5}$ \\
\hline
 & & & & & \\[-1em]
$\hat{N} = 60$ & 0.96935 & 0.96956 & 0.09655 & 0.09466 & $10^{-3}$ \\
\hline 
 & & & & & \\[-1em]
$\bar{N} = 60$ & 0.96911 & 0.96933 & 0.09736 & 0.09544 & $10^{-3}$ \\
\hline 
 & & & & & \\[-1em]
$\hat{N} = 60$ & 0.97451 & 0.97477 & 0.06796 & 0.06675 & $0.1$ \\
\hline 
 & & & & & \\[-1em]
$\bar{N} = 60$ & 0.97320 & 0.97348 & 0.07148 & 0.07013 & $0.1$ \\
\hline 
 & & & & & \\[-1em]
$\hat{N} = 60$ & 0.97482 & 0.97507 & 0.06716 & 0.06597 & $10$ \\
\hline
 & & & & & \\[-1em]
$\bar{N} = 60$ & 0.97276 & 0.97305 & 0.07264 & 0.07125 & $10$ \\
\hline
\end{tabular}
\caption{First and third order results for the observables of the nonminimal Coleman-Weinberg model considered in \cite{Kannike2016} for various values of the nonminimal coupling $\xi$ and for $\hat{N} = \bar{N} = 60$. We see that as $\xi$ grows so does the difference between the observables, depending on which definition for the $e$-folds we use.}
\label{table:E-vs-J-efolds}
\end{center}
\end{table}

\begin{figure}
\centering
\includegraphics[width=11.5cm]{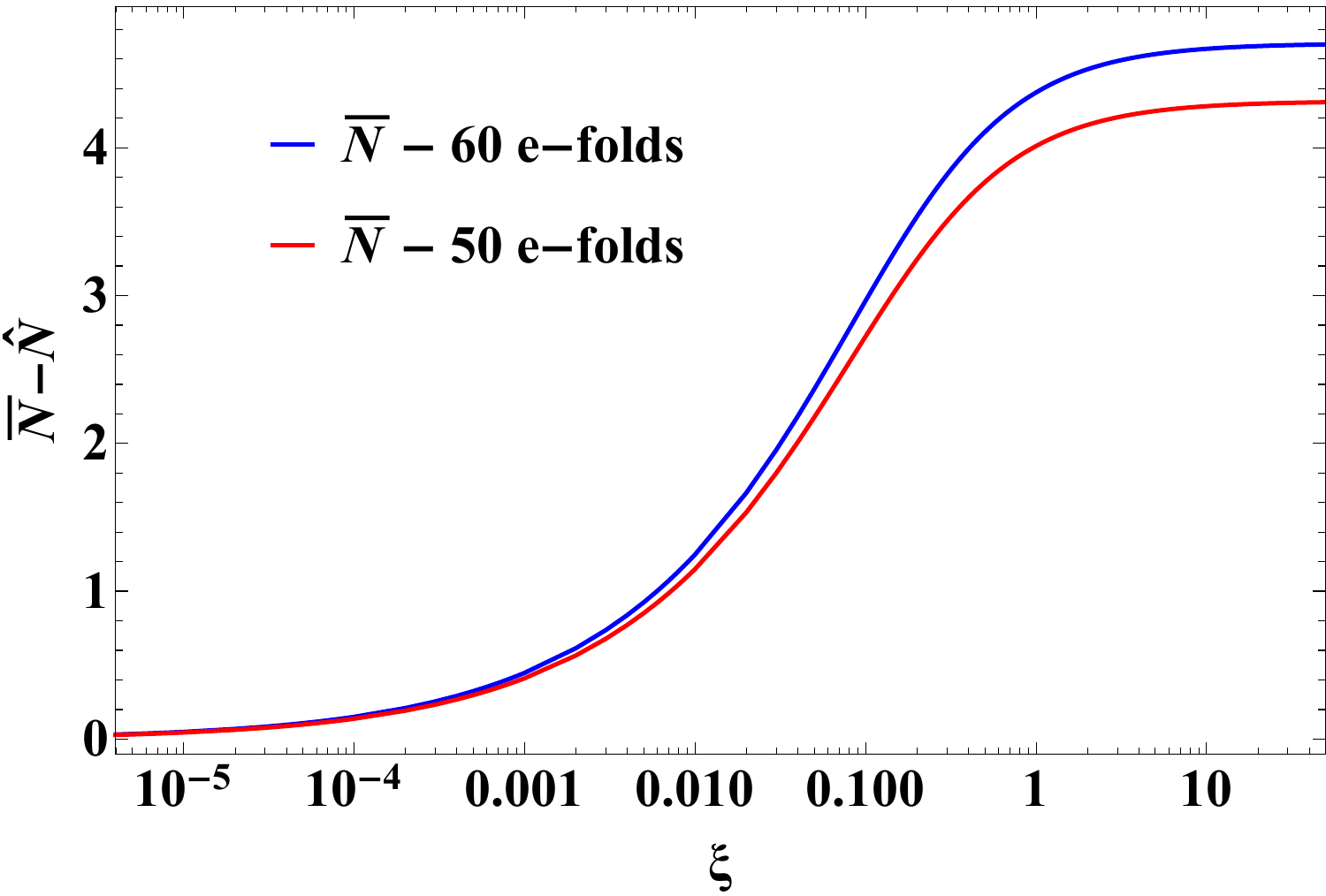}
\caption{The difference between the JF ($\bar{N}$) and the EF ($\hat{N}$) number of $e$-folds as a function of the nonminimal coupling $\xi$ for $\hat{N} = 60$ (top curve) and $\hat{N} = 50$ (bottom curve). We see that as $\xi$ grows we need more $e$-folds in the Jordan frame for the same inflaton field excursion.}
\label{fig:DeltaNFinal}
\end{figure}

Another way to illustrate the disparity between the two definitions for the $e$-folds is to examine how the same field excursion affects the number of $e$-folds itself. In Fig.~\ref{fig:DeltaNFinal}, for a wide range of values of $\xi$, we calculate the invariant $\Iphi^{\rm HC}$ for which $\hat{N} = 50$ and $\hat{N} = 60$. Then, for the same value of $\Iphi$ we calculate the corresponding JF $e$-folds $\bar{N}$ and plot the difference with the EF $e$-folds $\hat{N}$. One can see that, as expected, the difference asymptotes to zero for $\xi \rightarrow 0$ due to the vanishing second term in \eqref{Jefolds}. On the other hand, as $\xi$ grows so does the difference $\bar{N} - \hat{N}$ until it reaches a value of about $4.3$ $e$-folds for $\hat{N} = 50$ and $4.7$ $e$-folds for $\hat{N} = 60$. Note that for $\xi \gtrsim 10$ the difference stops growing since the model has reached the linear inflation attractor. We perceive the JF definition for the number of e-folds as the fundamental one since it is composed of all three invariants \eqref{inv_m}--\eqref{inv_phi} and also accommodates the EF definition.

\subsection{Taylor vs Pad\'{e}}

Let us also examine how the end-of-inflation condition affects the observables. Inflation ends \textit{exactly} at $\epsilon_H = 1$. Most authors usually adopt the slow-roll approximation and consider the relation between $\epsilon_H$ and the PSRPs at first order in the Taylor expansion and solve
\be 
\epsilon_H^{(\rm I)} = \epsilon_V = 1
\label{epsH-Taylor1}
\ee
in order to obtain the inflaton field value at the end of inflation. In our case, since we have obtained $n_S$ and $r$ at third order in the PSRPs, it would seem prudent to also approximate $\epsilon_H$ in the definition of $e$-folds with the third order Taylor expansion and solve 
\be 
\epsilon_H^{(\rm III)} =  \eps_V - \frac{4}{3} \eps_V^2+\frac{2}{3} \eps_V \eta_V+\frac{32}{9}\eps_V^3+\frac{5}{9} \eps_V \eta_V^2-\frac{10}{3} \eps_V^2 \eta_V+\frac{2}{9} \eps_V \zeta_V^2 = 1
\label{epsH-Taylor3}
\ee
in order to obtain $\Iphi^{\rm end}$. Nevertheless, even though the third order Taylor expansion is a very good approximation around the time of horizon crossing when the slow-roll parameters are small, the same does not hold near the end of inflation when $\epsilon_V$ and $\eta_V$ become of order one since the third order expansion actually blows up and thus fails to accurately describe the entirety of the inflationary epoch. A more accurate option, as pointed out in \cite{Liddle1994}, is to consider a Pad\'{e} approximation for $\epsilon_H$. The $\left[ 1/1 \right]$ Pad\'{e} approximant is given by
\be 
\epsilon_H^{\left[ 1/1 \right]} = \frac{\epsilon_V}{1 + \frac{4}{3} \epsilon_V - \frac{2}{3} \eta_V} ,
\label{epsH-Pade11}
\ee
while the $\left[ 2/2 \right]$ approximant has the form
\be
\begin{split}
\epsilon_H^{\left[ \rm{2/2} \right]} = & \, \frac{\epsilon_V+\frac{17}{4} \epsilon_V^2-\frac{5}{3} \epsilon_V \eta_V}{1+\frac{67}{12}\epsilon_V-\frac{7}{3}\eta_V-\frac{7}{2} \epsilon_V \eta_V+\frac{35}{9}\epsilon_V^2+\eta_V^2-\frac{2}{9}\zeta_V^2}
\\
& +\frac{2}{27} \eps_V \rho_V^3-\frac{1}{54} \eps_V^3 \eta_V+\frac{35}{108} \eps_V^2 \eta_V^2-\frac{13}{54} \eps_V^2 \zeta_V^2-\frac{1}{9} \eps_V \eta_V^3 .
\end{split}
\label{epsH-Pade22}
\ee

In Table \ref{table:Taylor-vs-Pade} we present the results for $n_S$ and $r$ for $\xi = 10^{-5}$, $\xi = 0.1$ and $\hat{N} = 50$ having employed the four end-of-inflation conditions for $\Iphi^{\rm end}$ described above and the corresponding expressions \eqref{epsH-Taylor1} - \eqref{epsH-Pade22} for $\epsilon_H$ in the $e$-folds integral. We find that the difference between the four methods is small for $n_S$ but larger for $r$ which has a greater dependence on $\epsilon_H$. The largest difference for $r$ between the methods occurs for small $\xi$ since its value is sizeable ($r \simeq 0.15$) and a small change in the value of $\Iphi^{\rm end}$ affects it noticeably. In any case, the differences between the end-of-inflation methods on $n_S$ and $r$ are comparable to the differences between the first and third order results.

\begin{table}
\begin{center}
\begin{tabular}{| c | c | c | c | c | c |}
\hline
 & & & & & \\[-1em]
$\mathbf{\hat{N} = 50}$ & $n_S^{(\rm I)}$ & $n_S^{(\rm III)}$ & $r^{(\rm I)}$ & $r^{(\rm III)}$ & $\xi$ \\
\hline 
 & & & & & \\[-1em] 
end: $\epsilon_H^{(\rm I)} = 1$ & 0.96078 & 0.96092 & 0.15238 & 0.14914 & $10^{-5}$ \\
\hline
 & & & & & \\[-1em]
end: $\epsilon_H^{\left[ 1/1 \right]} = 1$ & 0.95979 & 0.95994 & 0.15626 & 0.15285 & $10^{-5}$ \\
\hline 
 & & & & & \\[-1em]
end: $\epsilon_H^{(\rm III)} = 1$ & 0.96032 & 0.96047 & 0.15417 & 0.15085 & $10^{-5}$ \\
\hline
 & & & & & \\[-1em]
end: $\epsilon_H^{\left[ 2/2 \right]} = 1$ & 0.96019 & 0.96034 & 0.15468 & 0.15134 & $10^{-5}$ \\
\hline
 & & & & & \\[-1em]
end: $\epsilon_H^{(\rm I)} = 1$ & 0.96955 & 0.96991 & 0.08121 & 0.07948 & $0.1$ \\
\hline  
 & & & & & \\[-1em]
end: $\epsilon_H^{\left[ 1/1 \right]} = 1$ & 0.96870 & 0.96908 & 0.08348 & 0.08165 & $0.1$ \\
\hline 
 & & & & & \\[-1em]
end: $\epsilon_H^{(\rm III)} = 1$ & 0.96922 & 0.96959 & 0.08208 & 0.08031 & $0.1$ \\
\hline
 & & & & & \\[-1em]
end: $\epsilon_H^{\left[ 2/2 \right]} = 1$ & 0.96909 & 0.96946 & 0.08244 & 0.08066 & $0.1$ \\
\hline
\end{tabular}
\caption{First and third order results for the observables of the model \cite{Kannike2016} for two values of the nonminimal coupling $\xi$ and for $\hat{N} = 50$ using the four end-of-inflation conditions described in the text. We see that the differences are small albeit comparable to the differences between the first and third order results.}
\label{table:Taylor-vs-Pade}
\end{center}
\end{table}

\section{Summary and discussion}
\label{sec:Conclusions}

In the first part of this work we briefly reviewed the frame and reparametrization invariant formalism of scalar-tensor theories developed in \cite{Jaerv2015, Jaerv2015a, Kuusk2016, Kuusk2016a, Jaerv2017}. This formalism proves to be useful for inflation since it allows us to classify various models based on their invariant potentials. Therefore, it becomes transparent why theories with very different physical motivations yield similar predictions for the inflationary observables. 

Motivated by the imminent advancement in the sensitivity of the experiments, we then calculated the tensor and scalar spectral indices as well as the tensor-to-scalar ratio up to third order in the HSRPs in both the Einstein and Jordan frames employing the Green's function method introduced in \cite{Gong2001}. After this, utilizing the relation between the HSRPs in the two frames, we showed the equivalence of the frames. By construction, the Green's function method is valid to arbitrary order in the slow-roll expansion. Therefore, we expect the equivalence to hold up to any order. In addition, since the HSRPs are related to the PSRPs, we expressed the spectral indices and the ratio in terms of the PSRPs which are manifestly invariant.

Nevertheless, since the definition of the number of $e$-folds is different in the two frames, this can result to different predictions for the observables. We demonstrated this difference by considering the nonminimally coupled Coleman-Weinberg model examined in \cite{Kannike2016} and saw that as the nonminimal coupling grows so does the difference in the predictions. Such a difference can in fact be larger the differences between the first and third order results and will be detectable by the planned future experiments. \textit{We regard the Jordan frame definition for the number of e-folds \eqref{Jefolds} as \textit{the} fundamental one since it can be expressed in terms of all the principal invariants and also includes the Einstein definition}. Furthermore, we examined how various end-of-inflation conditions affect the inflationary observables. We found that the differences between the methods are comparable to the differences between the first and third order results. 

The above discussion proves that with the advent of precision experiments, care must be taken when analyzing a given inflationary model since the underlying methods and assumptions used may play an instrumental role in determining the viability of said model.

\section*{Acknowledgments}
T.P. would like to thank the Alexander S. Onassis Public Benefit Foundation for financial support.


\section*{Appendixes}

\begin{appendices}

\section{From Hubble to potential slow-roll parameters}
\label{app1:HSRPs-to-PSRPs}

The HSRPs are related to the PSRPs up to third order in the Taylor expansion via the following expressions \cite{Liddle1994}:
\begin{eqnarray}
\eps_H &=& \eps_V-\frac{4}{3} \eps_V^2+\frac{2}{3} \eps_V \eta_V+\frac{32}{9}\eps_V^3+\frac{5}{9} \eps_V \eta_V^2-\frac{10}{3} \eps_V^2 \eta_V+\frac{2}{9} \eps_V \zeta_V^2 ,\\
\eta_H &=& \eta_V -\eps_V+\frac{8}{3} \eps_V^2+\frac{1}{3} \eta_V^2-\frac{8}{3}\eps_V \eta_V +\frac{1}{3} \zeta_V^2 -12 \eps^3_V +\frac{2}{9} \eta_V^3+16 \eps_V^2 \eta_V \nonumber \\&& -\frac{46}{9} \eps_V \eta_V^2 -\frac{17}{9} \eps_V \zeta_V^2 +\frac{2}{3} \eta_V \zeta^2_V+\frac{1}{9} \rho_V^3 ,\\
\zeta_H^2 &=& \zeta_V^2-3 \eps_V \eta_V+3 \eps_V^2 -20 \eps_V^3+26 \eps_V^2 \eta_V-7 \eps_V \eta_V^2-\frac{13}{3} \eps_V \zeta_V^2+\frac{4}{3} \eta_V \zeta_V^2+\frac{1}{3} \rho_V^3 ,\\
\rho_H^3 &=& \rho_V^3 -3 \eps_V \eta_V^2+18 \eps_V^2 \eta_V -15 \eps_V^3-4 \eps_V \zeta_V^2 .
\end{eqnarray}

\section{Runnings of the spectral indices}
\label{app2:runnings}

The runnings of the tensor and scalar spectral indices up to third order in the HSRPs are given in the JF by
\ba 
\frac{\dd \bar{n}_T}{\dd \ln k} & = & - 2 \bk_0 \bk_1 - 6 \bk_0 \bk_1 \bl_0 - 4 \bk_0 \bl_0 \bl_1 - 6 \bk_0^2 \bk_1 + \left( 2 \alpha - 2 \right) \left( \bk_0 \bk_1^2 + \bk_0 \bk_1 \bk_1 \right) ,
\\
\frac{\dd \bar{n}_S}{\dd \ln k} & = & - 2 \bk_0 \bk_1 - \bk_1 \bk_2 - 6 \bk_0 \bk_1 \bl_0 - 4 \bk_0 \bl_0 \bl_1 - \bk_1 \bl_0 \bl_1 - 2 \bk_1 \bk_2 \bl_0 - 2 \bl_0 \bl_1 \bl_2 \nonumber \\ 
&& - 2 \bl_0 \bl_1^2 - 6 \bk_0^2 \bk_1 + \left( 2 \alpha - 3 \right) \bk_0 \bk_1^2 + \left( 2 \alpha - 4 \right) \bk_0 \bk_1 \bk_2 + \alpha \left( \bk_1 \bk_2^2 + \bk_1 \bk_2 \bk_3 \right) ,
\ea
while in the EF the runnings have the form
\ba 
\frac{\dd \hat{n}_T}{\dd \ln k} & = & - 2 \hk_0 \hk_1 - 6 \hk_0^2 \hk_1 + \left( 2 \alpha - 2 \right) \left( \hk_0 \hk_1^2 + \hk_0 \hk_1 \hk_1 \right) ,
\\ 
\frac{\dd \hat{n}_S}{\dd \ln k} & = &  - 2 \hk_0 \hk_1 - \hk_1 \hk_2 - 6 \hk_0^2 \hk_1 + \left( 2 \alpha - 3 \right) \hk_0 \hk_1^2 + \left( 2 \alpha - 4 \right) \hk_0 \hk_1 \hk_2 \nonumber \\
&&  + \alpha \left( \hk_1 \hk_2^2 + \hk_1 \hk_2 \hk_3 \right) .
\ea
Again, plugging \eqref{hk0-to-bk0} - \eqref{tripleh-to-tripleb} into the EF expressions, one can see that the expressions for the runnings of the spectral indices in the two frames coincide. Finally, the runnings of the spectral indices can be written in terms of the PSRPs as
\ba 
\frac{\dd n_T}{\dd \ln k} & = &  - 8 \epsilon_V^2 + 4 \epsilon_V \eta_V + \left( 52 \alpha - \frac{148}{3} \right) \epsilon_V^3 - \left( 50 \alpha - 38 \right) \epsilon_V^2 \eta_V \nonumber \\
&&+ \left( 16 \alpha - 12 \right) \epsilon_V \eta_V^2 + \left( 4 \alpha - \frac{8}{3} \right) \epsilon_V \zeta_V^2 ,\\
\frac{\dd n_S}{\dd \ln k} & = & - 24 \epsilon_V^2 + 16 \epsilon_V \eta_V - 2 \zeta_V^2 + \left( 180 \alpha - \frac{104}{3} \right) \epsilon_V^3 - \left( 180 \alpha + \frac{4}{3} \right) \epsilon_V^2 \eta_V \nonumber \\ 
&& + \left( 32 \alpha + 12 \right) \epsilon_V \eta_V^2 + \left( 24 \alpha + 4 \right) \epsilon_V \zeta_V^2 - \left( 2 \alpha - \frac{8}{3} \right) \eta_V \zeta_V^2 - \left( 2 \alpha + \frac{2}{3} \right) \rho_V^3 .
\ea

\section{Equation of motion in terms of $e$-folds}
\label{app3:efolds}
We can rewrite the equation of motion for the invariant $\Iphi$ as a nonlinear second order differential equation with respect to the number of $e$-folds. In the Einstein frame we have
\beq
\frac{\dd^2 \Iphi }{\dd \hat{N}^2}+3\frac{\dd  \Iphi }{\dd \hat{N}}-\left( \frac{\dd  \Iphi }{\dd \hat{N}}\right)^3+\left[ 1-\frac{1}{3}\left( \frac{\dd  \Iphi }{\dd \hat{N}} \right)^2 \right] 3\sqrt{\eps_V} = 0 ,
\label{Eefolds-EOM}
\eeq
while in the Jordan frame the equation of motion can be brought to the following form:
\be 
\begin{split}
\frac{\dd^2 \Iphi}{\dd \bar{N}^2}+3 \frac{\dd\Iphi}{\dd \bar{N}}&+\frac{\dd\Iphi}{\dd \bar{N}} \left[ 1-\frac{1}{2} \frac{\dd \ln{\Inm}}{\dd \bar{N}} \right]^{-1} \left[ -\frac{1}{2}\frac{\dd \ln{\Inm}}{\dd \bar{N}}+\frac{1}{4}\left( \frac{\dd \ln{\Inm}}{\dd \bar{N}} \right)^2 - \left( \frac{\dd\Iphi}{\dd \bar{N}} \right)^2+\frac{1}{2} \frac{\dd^2 \ln{\Inm}}{\dd \bar{N}^2} \right]
\\
&-\frac{\dd \ln{\Inm}}{\dd \bar{N}}\frac{\dd\Iphi}{\dd \bar{N}}+\left[ 1 + \frac{1}{4} \left(\frac{\dd \ln{\Inm}}{\dd \bar{N}} \right)^2-\frac{\dd \ln{\Inm}}{\dd \bar{N}}-\frac{1}{3}\left( \frac{\dd\Iphi}{\dd \bar{N}}\right)^2 \right] 3\sqrt{\eps_V} = 0 .
\end{split}
\label{Jefolds-EOM}
\ee
By numerically solving these equations we can obtain the invariant field as a function of the number of e-folds in the two frames. Of course, in the case with minimal coupling we have $\frac{\dd \ln{\Inm}}{\dd \bar{N}} = \frac{\dd^2 \ln{\Inm}}{\dd \bar{N}^2} = 0$ and $\bar{N} = \hat{N}$, which means that \eqref{Jefolds-EOM} reduces to \eqref{Eefolds-EOM}.

\end{appendices}

\bibliography{References}{}
\bibliographystyle{utphys}

\end{document}